\documentclass[prd,nofootinbib,preprintnumbers,twocolumn,showpacs,floatfix,superscriptaddress]{revtex4-1}

\usepackage{amsmath}
\usepackage{amssymb}
\usepackage{graphicx}
\usepackage{color}
\usepackage{hyperref}

\newcommand{\bitem}{\begin{itemize}}
\newcommand{\eitem}{\end{itemize}}
\newcommand{\bwt}{\begin{widetext}}
\newcommand{\ewt}{\end{widetext}}
\newcommand{\beq}{\begin{equation}}
\newcommand{\eeq}{\end{equation}}

\newcommand{\bdm}{\begin{displaymath}}
\newcommand{\edm}{\end{displaymath}}
\newcommand{\bea}{\begin{eqnarray}}
\newcommand{\eea}{\end{eqnarray}}

\newcommand{\nn}{\nonumber}

\def\eq#1{{Eq.~(\ref{#1})}}
\def\eqs#1#2{{Eqs.~(\ref{#1})--(\ref{#2})}}
\def\Eq#1{{Eq.~\ref{#1}}}

\def\fig#1{{Fig.~\ref{#1}}}

\def\Table#1{{Table~\ref{#1}}}

\def\sect#1{{Sect.~\ref{#1}}}

\def\app#1{{Appendix~\ref{#1}}}

\def\vev#1{\left\langle #1 \right\rangle}

\def\abs#1{\left| #1\right|}

\def\Tr{\!\mathop{\rm Tr}}
\def\det{\mbox{det}\,}

\newcommand\GeV{\text{GeV}}

\begin{document}

\title{Massive neutrinos and invisible axion minimally connected}
\author{Stefano Bertolini}
\email{stefano.bertolini@sissa.it}
\affiliation{INFN, Sezione di Trieste, SISSA,
Via Bonomea 265, 34136 Trieste, Italy\\[1ex]}

\author{Luca Di Luzio}
\email{luca.di.luzio@ge.infn.it}
\affiliation{Dipartimento di Fisica, Universit\`a di Genova and INFN, Sezione di Genova,
Via Dodecaneso 33, 16159 Genova, Italy\\[1ex]}

\author{Helena Kole\v{s}ov\'a}
\email{helena.kolesova@fjfi.cvut.cz}
\affiliation{Faculty of Nuclear Sciences and Physical Engineering,
Czech Technical University in Prague,
Brehov\'a 7, 115 19 Praha 1, Czech Republic\\[1ex]}
\affiliation{Institute of Particle and Nuclear Physics,
Faculty of Mathematics and Physics,
Charles University in Prague, V Hole\v{s}ovi\v{c}k\'ach 2,
180 00 Praha 8, Czech Republic\\[1ex]}
\affiliation{SISSA/ISAS,
Via Bonomea 265, 34136 Trieste, Italy\\[1ex]}

\author{Michal Malinsk\'y}
\email{malinsky@ipnp.troja.mff.cuni.cz}
\affiliation{Institute of Particle and Nuclear Physics,
Faculty of Mathematics and Physics,
Charles University in Prague, V Hole\v{s}ovi\v{c}k\'ach 2,
180 00 Praha 8, Czech Republic\\[1ex]}


\begin{abstract}

\noindent
We survey a few minimal scalar extensions of the standard electroweak
model that provide a simple setup for massive neutrinos in
connection with an invisible axion. The presence of a
chiral $U(1)$ \`a la Peccei-Quinn drives the pattern of {Majorana} neutrino
masses while providing a dynamical solution to the strong CP problem and an axion as a dark matter candidate.
We paradigmatically apply such a renormalizable framework to type-II {seesaw} and to two viable models for neutrino oscillations where the neutrino {masses arise} at one and two loops, respectively. We comment on the naturalness of the effective setups as well as on their implications for vacuum stability and electroweak baryogenesis. 

\end{abstract}

\maketitle

\section{Introduction}

\noindent

\begin{figure*}[t]
\centerline{
\includegraphics[width=1.8\columnwidth]{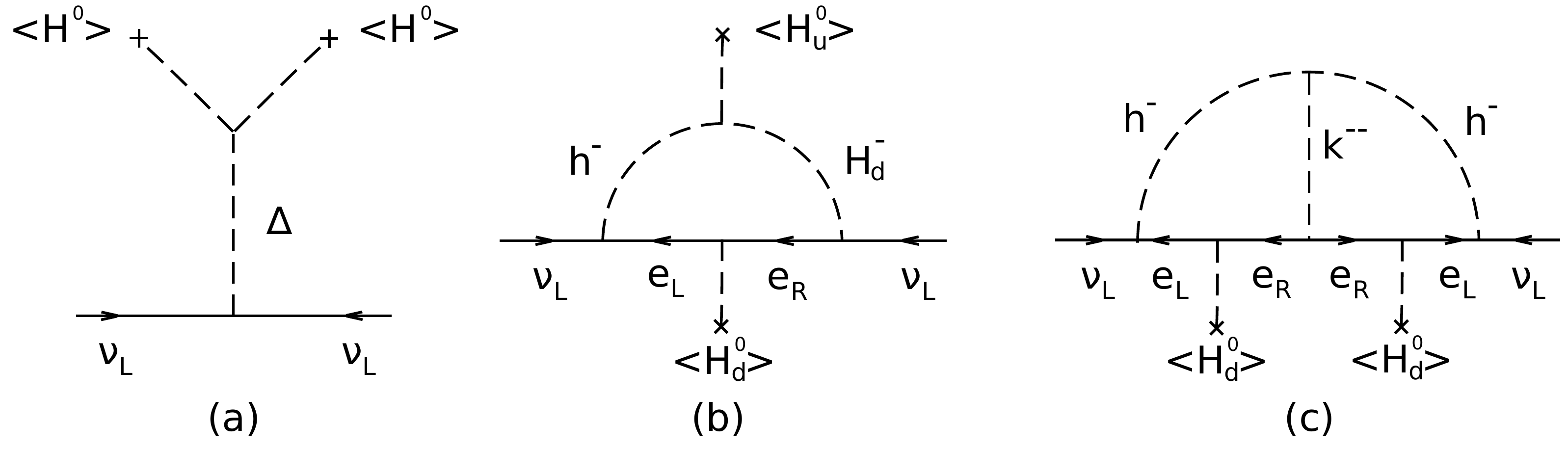}
}
\vspace*{-2ex}
\caption{Sample diagrams leading to the $\Delta L =2$ Weinberg operator at the tree level (a), one loop (b) and two loops (c) {in the type-II seesaw, Zee and Zee-Babu models, respectively.}}
\label{fig:numassdiag}
\end{figure*}

The first LHC run has led to the discovery of a scalar particle that looks much {like} the Higgs boson of the $SU(2)_L\otimes U(1)_Y$ electroweak standard model (SM). The raising limits on exotic physics scales {set a challenge to the popular issue of naturalness}~\cite{Farina:2013mla}, a paradigm that has guided much of the beyond the SM modelling in the last decades. This notwithstanding, neutrino oscillations and dark matter {call} for physics beyond the standard scenario. Baryon asymmetry calls for it as well while electroweak vacuum stability may not be an issue in minimally extended scenarios~\cite{EliasMiro:2012ay}.
We aim at discussing a class of minimal extensions of the SM that account for the aforementioned open issues. To this {end} we choose to maintain the fermionic SM content as it stands and consider only extensions of the scalar sector. Advantages of this choice will be clear in the following. {According to that,} the only tree-level realization of the dimension-5 Weinberg operator $(LLHH)/M$ for Majorana neutrino masses is via the mediation of an $SU(2)_L$ scalar triplet of hypercharge one. This is commonly known as the type-II {seesaw}~\cite{Schechter:1980gr,Cheng:1980qt,Lazarides:1980nt,Mohapatra:1980yp,Wetterich:1981bx}, Fig.~1a. 

At the radiative level an elegant and simple realization {of the same} was provided long ago by Zee~\cite{Zee:1980ai}; the Weinberg operator is {there} obtained at one loop from the dimension-7 effective operator  $(LLLe^c H)/M^3$~\cite{Ma:1998dn,Babu:2001ex,Bonnet:2012kz} when $L$ and $e^c$ are connected by the $H$ Yukawa coupling ({giving rise to a} chiral suppression), as shown in Fig.~1b.
The model requires one additional weak doublet and a weak scalar singlet of hypercharge one. In order to avoid Higgs mediated flavor changing neutral currents a $Z_2$ symmetry is called for~\cite{Wolfenstein:1980sy}. Such a model, however, is not consistent with the neutrino oscillation data~\cite{Koide:2001xy,Frampton:2001eu,He:2003ih}. Recently, Babu and Julio (BJ)~\cite{Babu:2013pma} presented a variant of the Zee model with a $Z_4$ discrete family symmetry that restores consistency with the observed neutrino mixing pattern. The model yields an inverted {neutrino mass} hierarchy and is highly predictive for neutrinoless double beta decay and lepton flavor violation (LFV). 

{At two loops, a popular} realization of the Weinberg operator is given by the Zee-Babu (ZB) model~\cite{Zee:1985id,Babu:1988ki}. {In this setting,} the neutrino mass matrix is obtained by dressing the dimension-9 effective operator $(LLLe^cLe^c)/M^5$~\cite{Babu:2001ex,Sierra:2014rxa} and it requires two weak scalar singlets with hypercharge one and two, respectively (Fig.~1c). It is a very simple extension of the SM that leads to calculable neutrino masses and mixings in agreement with all present oscillation and lepton flavor phenomenology (for a recent reappraisal see~\cite{Herrero-Garcia:2014hfa,Schmidt:2014zoa}).

Our interest is to discuss simple renormalizable  extensions of the
standard scenario that are effective at the TeV scale {and lead to testable signals} at the available energy and foreseen intensity
facilities. There is one inherent large scale involved that is linked
to the presence of a spontaneously broken Peccei-Quinn (PQ)
symmetry~\cite{Peccei:1977hh,Peccei:1977ur} and the related
axion~\cite{Weinberg:1977ma,Wilczek:1977pj}. As we shall discuss, it
is noteworthy that the presence of such a large scale (above $ 10^{9}$ GeV) does not endanger the radiative stability of the setup. While the anomalous $U(1)_{PQ}$ gives an elegant solution to the so-called strong CP problem in QCD~\cite{Weinberg:1975ui,'tHooft:1976fv,Callan:1976je,Jackiw:1976pf,'tHooft:1986nc}, the axion provides a viable dark matter candidate (see~\cite{Kim:2008hd} for a recent review).
We find {it} appealing and intriguing that a simple renormalizable framework can be conceived where the origin of neutrino masses and the solution of the strong CP problem are fundamentally related and where the requirement of naturalness and stability of the scalar sector is tightly linked to the light neutrino scale.
   
The idea of connecting massive neutrinos with the presence of a spontaneously broken $U(1)_{PQ}$ comes a long way~\cite{Mohapatra:1982tc,Shafi:1984ek,Langacker:1986rj,Shin:1987xc,He:1988dm,Geng:1988nc,Berezhiani:1989fp,Bertolini:1990vz,Arason:1990sg,Ma:2001ac,Dias:2005dn,Ma:2011rp,Chen:2012baa,Park:2014qha,Dias:2014osa,Ahn:2014gva,Ma:2014yka}.  Considering only scalar extensions of the SM a simple setup based on the Zee model for radiative neutrino masses was discussed in~\cite{Bertolini:1990vz,Arason:1990sg}. The model features a Dine, Fischler, Srednicki,
and Zhitnitsky (DFSZ)\footnote{No extension of the matter sector is needed at variance with the class of invisible axion models proposed by Kim, Shifman, Vainshtein and Zakharov~\cite{Kim:1979if,Shifman:1979if} (KSVZ) that feature a vector-like quark.} invisible axion~\cite{Zhitnitsky:1980tq,Dine:1981rt}, with a tiny coupling to neutrinos. The need for two different Higgs doublets and the role of the related $Z_2$ symmetry are there a free benefit of the minimal implementation of the anomalous PQ symmetry. Two additional neutral and two singly charged scalars remain naturally light (TeV scale). In spite of the presence of the large PQ scale the model is shown to exhibit 
a radiatively stable hierarchy. In all analogy with the Zee model, a simple {Majorana} neutrino mass matrix with vanishing diagonal entries arises at one-loop, whose structure is determined by three parameters. As already mentioned such a structure is shown to exhibit nearly bi-maximal mixing and it is ruled out by oscillation data.

In this paper we show how this setup can {work in general}. We discuss three explicit viable schemes: the paradigmatic low-scale    
type-II {seesaw} (TII), the one-loop BJ model and the two-loop ZB model. In the extended BJ model a lepton-family-dependent PQ symmetry plays the role of the original $Z_4$ symmetry. In all cases one obtains a DFSZ invisible axion with a tiny coupling to neutrinos. In the BJ case the axion exhibits flavour violating couplings to the leptons of the same size of the diagonal ones. 
Such flavour violating couplings are not directly constrained by astrophysical processes 
and future laboratory tests of LFV might even provide competitive bounds on the PQ scale \cite{Jaeckel:2013uva}.   
In addition to {a} heavy neutral scalar (mainly) singlet the physical scalar spectrum exhibits in the three models two singly-charged and two additional neutral states. In the case of TII and ZB a doubly charged scalar is present as well with a distinctive role in 
LFV phenomenology. 

Stability of the scalar sector demands tiny interactions {between} the PQ heavy state and the remaining scalars. {Due to an enhanced symmetry in the vanishing interaction limit,} the smallness of the relevant couplings is preserved at higher orders. {Remarkably, such a setup allows for naturally light neutrinos together with a rich scalar spectrum at the TeV scale} . The possible presence of {an exotic TeV-scale} scalar sector is not yet excluded by collider searches and it is {among the priorities} in the coming years.

A fringe benefit of such an extension of the standard scalar sector is to improve the electroweak vacuum stability. On the other hand, the sizable interactions among the ``light'' scalar states open a possibility for the realization of a first-order electroweak phase transition. This is one of the requirements for electroweak baryogenesis~\cite{Morrissey:2012db}. However, no new sources of CP violation arise from the minimal scalar sectors featured in the considered setups. We shall comment on the possibility of addressing baryogenesis within such a framework. 

In the next three sections we detail the extended TII, BJ and ZB setups {and  discuss their generic features and shortcomings} in \sect{discussion}.

\section{PQ extended type-II seesaw}\label{sec:typeIIseesaw}

On top of the usual SM field content, the scalar sector of the PQ extended Type-II {seesaw} 
model features two Higgs doublets, 
an isospin triplet with hypercharge one and a SM singlet (cf.~\Table{fctypeII}). 
The PQ charge assignments are displayed in \Table{fctypeII}, 
where the presence of Yukawa interactions for quarks is already taken into account. 
Recall that the PQ current is axial, thus proportional to the difference between the charges of the left- and right-handed 
(colored) fermions. Hence, without loss of generality, we can always set $X_q = 0$. In this way, 
the color anomaly of the PQ current is proportional to $X_u + X_d$ (see, e.g.,~\cite{Srednicki:1985xd}).

\begin{table}[thbp]
  \centering
  \begin{tabular}{@{} |c|c|c|c|c|c| @{}}
 \hline
    Field & Spin & $SU(3)_C$ & $SU(2)_L$ & $U(1)_Y$ & $U(1)_{PQ}$ \\ 
 \hline
    $q_L$ & $\frac{1}{2}$ & 3  & 2 & $+\frac{1}{6}$ & 0\\ 
    $u_R$ & $\frac{1}{2}$ & 3  & 1 & $+\frac{2}{3}$ & $X_u$ \\ 
    $d_R$ & $\frac{1}{2}$ & 3  & 1 & $-\frac{1}{3}$ & $X_d$ \\ 
    $\ell_L$ & $\frac{1}{2}$ & 1  & 2 & $-\frac{1}{2}$ & $X_\ell$ \\ 
    $e_R$ & $\frac{1}{2}$ & 1  & 1 & $-1$ & $X_e$ \\
    $H_u$ & 0 & 1 & 2 & $-\frac{1}{2}$ & $-X_u$ \\
    $H_d$ & 0 &  1  & 2 & $+\frac{1}{2}$ & $-X_d$ \\
    $\Delta$ & 0 &  1  & 3 & +1 & $X_\Delta$ \\
    $\sigma$ & 0 &  1  & 1 & 0 & $X_\sigma$ \\
 \hline
  \end{tabular}
  \caption{\label{fctypeII} Field content and charge assignment of the PQ extended Type-II seesaw model.}
\end{table}

\subsection{Lagrangian}

The only two sectors which are sensitive to the assignment of the PQ charges are the Yukawa Lagrangian and the scalar potential that we discuss in turn.
The former reads
\begin{multline}
\label{Yukawa}
- \mathcal{L}^{\rm{TII}}_{Y} = 
Y_u \, \overline{q}_{L} u_{R} H_u 
+ Y_d \, \overline{q}_{L} d_{R} H_d 
+ Y_e \, \overline{\ell}_{L} e_{R} H_d \\ 
+ \tfrac{1}{2} Y_\Delta \, \ell_L^T C i \tau_2 \mathbf{\Delta} \ell_L
+ \rm{h.c.} \, , 
\end{multline}
where the flavour contractions are understood (e.g.~$Y_\Delta^T = Y_\Delta$),
$C$ is the charge conjugation matrix in {the} spinor space, and
\begin{equation}
\label{Qeigenstates}
\mathbf{\Delta} \equiv 
\frac{\vec{\tau} \cdot \vec{\Delta}}{\sqrt{2}}  
= 
\left(
\begin{array}{cc}
\frac{\Delta^+}{\sqrt{2}} & \Delta^{++} \\
\Delta^{0} & - \frac{\Delta^+}{\sqrt{2}}
\end{array}
\right)
\, . 
\end{equation}
In \eq{Qeigenstates}, $\vec{\tau} = (\tau_1, \tau_2, \tau_3)$ are the Pauli matrices and 
$\vec{\Delta} = (\Delta_1, \Delta_2, \Delta_3)$ 
are the $SU(2)_L$ components of the scalar triplet. 
The electric charge eigenstates are obtained by the action of $Q = T_3 + Y$ on \eq{Qeigenstates}.

The scalar potential can be written as~\cite{Bertolini:1990vz,Arhrib:2011uy}
\begin{align}
\label{TII-scalarpot}
V_{\rm{TII}} &= -\mu_1^2 \abs{H_u}^2 + \lambda_1 \abs{H_u}^4 -\mu_2^2 \abs{H_d}^2 + \lambda_2 \abs{H_d}^4 \nonumber \\
&+ \lambda_{12} \abs{H_u}^2 \abs{H_d}^2 + \lambda_{4} \abs{H_u^\dag H_d}^2 \nonumber \\
& - \mu_3^2 \abs{\sigma}^2 + \lambda_3  \abs{\sigma}^4 
+ \lambda_{13} \abs{\sigma}^2 \abs{H_u}^2
+ \lambda_{23} \abs{\sigma}^2 \abs{H_d}^2 \nn \\
& + \Tr (\mathbf{\Delta}^\dag \mathbf{\Delta}) \Big[\mu_\Delta^2 + \lambda_{\Delta1} \abs{H_u}^2 + \lambda_{\Delta2} \abs{H_d}^2 \nn \\
& + \lambda_{\Delta3} \abs{\sigma}^2 + \lambda_{\Delta4} \Tr (\mathbf{\Delta}^\dag \mathbf{\Delta}) \Big] \nonumber \\
& + \lambda_7 H_u^\dag \mathbf{\Delta} \mathbf{\Delta}^\dag H_u 
+ \lambda_8 H_d^\dag \mathbf{\Delta} \mathbf{\Delta}^\dag H_d
+ \lambda_9  \Tr ( \mathbf{\Delta}^\dag \mathbf{\Delta} )^2 \nn \\
& + \left( \lambda_5 \sigma^2 \tilde{H}^\dag_u H_d + \lambda_6 \sigma H_u^\dag \mathbf{\Delta}^\dag H_d + \text{h.c.} \right)
\, ,  
\end{align}
where we employed the notation $\tilde{H}_u = i \tau_2 H_u^*$.
Notice that terms like $\tilde{H}^\dag_u H_d \Tr (\mathbf{\Delta}^\dag \mathbf{\Delta})$ or 
$\tilde{H}^\dag_u \mathbf{\Delta} \mathbf{\Delta}^\dag H_d$ 
are not allowed since the QCD anomaly of the PQ current requires $X_u + X_d\neq 0$. 
Moreover, $H_{u,d}^\dag \left( \mathbf{\Delta}^\dag \mathbf{\Delta} + \mathbf{\Delta} \mathbf{\Delta}^\dag \right) H_{u,d} 
= \abs{H_{u,d}}^2 \Tr (\mathbf{\Delta}^\dag \mathbf{\Delta})$, 
so that only two invariants out of three are linearly independent. 

The terms 
$\lambda_5 \, \sigma^2 \tilde{H}^\dag_u H_d$ and $\lambda_6 \, \sigma H_u^\dag \mathbf{\Delta}^\dag H_d$ are needed in order to assign a non-vanishing PQ charge to the singlet $\sigma$ and to generate neutrino masses. Notice that the simultaneous presence of  $\lambda_5$, $\lambda_6$ and $Y_\Delta$ is needed to explicitly break lepton number.  If any of the couplings is missing, either lepton number is exact and neutrinos are massless or lepton number is spontaneously broken and the vacuum exhibits a majoron together with a Wilczek-Weinberg axion~\cite{Bertolini:1990vz}. As shown next, the potential in \eq{TII-scalarpot} corresponds to a unique PQ charge assignment that forbids among else the presence of trilinear interaction terms. The absence of cubic scalar interactions, which characterizes the three models here discussed, paves the way to their embedding in a classically scale invariant  setup dynamically broken a la Coleman-Weinberg. We shall comment on that in \sect{conclusions}.

Finally, the couplings $\lambda_5$ and $\lambda_6$ can be set real by two independent rephasings of the fields.

\subsection{PQ charges}
\label{PQchargesTII}

The invariants in \eq{Yukawa} and \eq{TII-scalarpot} enforce the following constraints on the PQ charges: 
\begin{align}
\label{PQeq1}
- X_\ell + X_e - X_d &= 0 \, , \\
\label{PQeq2}
2 X_\ell + X_\Delta &= 0 \, , \\
\label{PQeq3}
2 X_\sigma - X_u  - X_d &= 0 \, , \\
\label{PQeq4}
X_\sigma + X_u  - X_\Delta - X_d &= 0 \, .
\end{align}
Solving in terms of $X_u$ and $X_d$ we get: 
\begin{align}
\label{PQchargesXuXd}
& X_\ell = - \frac{3 X_u}{4} + \frac{X_d}{4} \, , \quad 
X_e = - \frac{3 X_u}{4}  + \frac{5 X_d}{4} \, , \nn \\
& X_\Delta = \frac{3 X_u}{2}  - \frac{X_d}{2} \, , \quad \ \
X_\sigma = \frac{X_u}{2}  + \frac{X_d}{2} \, . 
\end{align} 
Following \cite{Dine:1981rt,Bertolini:1990vz} we require the orthogonality of the hypercharge and axion {currents}. 
This leads to the relation
\begin{equation}
\label{orthogonality}
X_u v_u^2 = X_d v_d^2 \, , 
\end{equation}
where $v_u = \vev{H_u}$ and $v_d = \vev{H_d}$. Adding this condition to \eq{PQchargesXuXd}, we can determine all the 
PQ charges up to an overall normalization. We choose this normalization by the condition
\begin{equation}
X_\sigma = 1 \, .
\end{equation}
By defining $x \equiv v_u / v_d$ the remaining charges in \eq{PQchargesXuXd} read 
\begin{align}
\nonumber &X_u = \frac{2}{x^2 + 1}\,,& \quad  &X_d = \frac{2 x^2}{x^2 + 1}\, ,& \quad &X_\ell = \frac{x^2 - 3}{2 (x^2 + 1)}\, , \\
\label{PQchargesx}&X_e = \frac{5 x^2 - 3}{2 (x^2 + 1)}  \, ,& \quad &X_\Delta = \frac{3 - x^2}{x^2 + 1}\,.& & 
\end{align}

\subsection{Scalar spectrum}
\label{scalarspectrumTII}

To compute the scalar spectrum we expand the scalar fields around the chargeless and CP-conserving 
vacuum configuration
\begin{align}
\label{expHu}
H_u &= 
\left(
\begin{array}{c}
v_u + \frac{h^0_u + i \eta^0_u}{\sqrt{2}} \\
h^-_u
\end{array} 
\right) \, , \\ 
\label{expHd}
H_d &= 
\left(
\begin{array}{c}
h^+_d \\
v_d + \frac{h^0_d + i \eta^0_d}{\sqrt{2}} 
\end{array} 
\right) \, , \\
\label{expDelta}
\mathbf{\Delta} &= 
\left(
\begin{array}{cc}
\frac{\delta^+}{\sqrt{2}} & \delta^{++} \\
v_\Delta + \frac{\delta^0 + i \eta^0_\delta}{\sqrt{2}} & - \frac{\delta^+}{\sqrt{2}}
\end{array} 
\right) \, , \\
\label{expsigma}
\sigma &= V_\sigma + \frac{\sigma^0 + i \eta^0_\sigma}{\sqrt{2}} \, ,
\end{align}
with $v_u$, $v_d$, $v_\Delta$ and $V_\sigma$ denoting the relevant (real) vacuum expectation values (VEVs).~\footnote{While it is assumed that there exists a region of the scalar potential parameters for which the absolute minimum preserves the electric charge, 
it can be shown (see \sect{EWbaryogenesis}) that the potential of \eq{TII-scalarpot} does not lead to 
spontaneous CP violation.} 

The scalar spectrum of the model is detailed in Appendix \ref{scalarspectrumII} and the main features are discussed in 
\sect{discussion}. Here we just anticipate that the model features a DFSZ invisible axion, with a tiny coupling to neutrinos, and its SM singlet companion with a PQ scale mass. By invoking a technically natural ultraweak limit (see the discussion in \sect{Naturalness}) such a heavy scalar is sufficiently decoupled from all the other physical scalar states that are requested to live at the TeV scale thus {preserving the radiative stability of the light scalar spectrum}.
At the weak scale the model allows for a SM-like Higgs boson; {this, together with a brief account of the relevant phenomenological constraints, shall be discussed in \sect{scalarph}.}

\subsection{Neutrino masses}

\begin{figure}
\begin{center}
\includegraphics[width=.6\columnwidth]{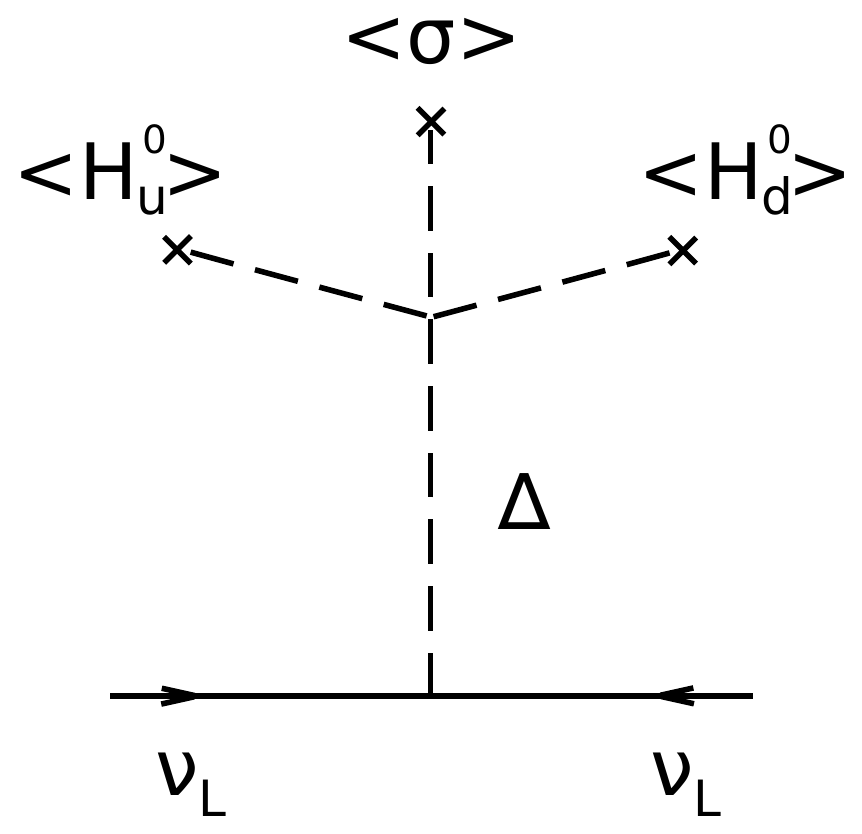}
\caption{The tree-level ``hug'' diagram responsible for the {Majorana} neutrino  mass in the PQ extended type-II {seesaw} model.}
\label{Fig:axseesawII}
\end{center}
\end{figure}

{In the TII model, the} neutrino masses are generated through the tree-level diagram in \fig{Fig:axseesawII}. 

Their expression is conveniently obtained by computing the (induced) VEV of the triplet.  
Let us hence consider the projection of the scalar potential along the neutral VEV components of \eqs{expHu}{expsigma}
\begin{multline}
\label{projectedV}
\vev{V_{\rm{TII}}} = \left(\mu_\Delta^2 + \lambda_{\Delta3} V_\sigma^2 
+ \lambda_{\text{$\Delta $1}} v_u^2
+ (\lambda _{\text{$\Delta $2}}
+ \lambda _8) v_d^2
\right) v_\Delta^2 \\
+ 2 \lambda_6 V_\sigma v_u v_d v_\Delta 
+ \mathcal{O} \left(  v_\Delta^4  \right) + v_\Delta \text{-indep.~terms} \, . 
\end{multline}
Given the phenomenological hierarchy $V_\sigma \gg v_{u,d} \gg v_\Delta$, the stationary condition with respect to $v_\Delta$ is well approximated by 
\begin{equation}
\label{stateqvD}
2 M_\Delta^2 v_\Delta 
+ 2 \lambda_6 V_\sigma v_u v_d \approx 0 \, ,
\end{equation}
where we defined the effective mass parameter 
\begin{equation}
M_\Delta^2 = \mu_\Delta^2 
+ \lambda_{\Delta3} V_\sigma^2 + \lambda_{\text{$\Delta $1}} v_u^2
+ (\lambda _{\text{$\Delta $2}}
+ \lambda _8) v_d^2 \, .
\end{equation}
In the decoupling limit $v_{u,d}/V_\sigma \rightarrow 0$, this
coincides with the triplet mass in the PQ-broken phase (cf.~\eq{M2triplet}). 
Hence, from \eq{stateqvD}, the induced VEV {of} $\mathbf{\Delta}$ reads
\begin{equation}
\label{inducedvev}
v_\Delta \approx 
\frac{\lambda_6 V_\sigma v_u v_d}{M_\Delta^2} \, .
\end{equation}
Since the triplet VEV breaks {the SM} custodial symmetry, it is bounded by {the} electroweak precision observables to 
$v_\Delta \lesssim 1$ GeV. {This, in turn,} implies the following bound on the scalar potential coupling {$\lambda_{6}$:}
\begin{equation}
\lambda_6 \lesssim 10^{-9} \left( \frac{10^9 \ \rm{GeV}}{V_\sigma} \right) \left( \frac{M^2_\Delta}{v_u v_d} \right) \, .
\end{equation}
Finally, from the Yukawa Lagrangian in \eq{Yukawa} we obtain
\begin{equation}
\label{neutrinomasses}
M_\nu^{\rm{TII}} = Y_\Delta v_\Delta \approx \frac{Y_\Delta \lambda_6 V_\sigma v_u v_d}{M^2_\Delta} \, ,
\end{equation}
as diagrammatically represented by the graph in \fig{Fig:axseesawII}, 
and the bound on the heaviest neutrino $m_{\nu_3} \lesssim 1$ eV translates into the constraint 
\begin{equation}
\lambda_6 Y_\Delta \lesssim 10^{-18} \left( \frac{10^9 \ \rm{GeV}}{V_\sigma} \right) \left( \frac{M^2_\Delta}{v_u v_d} \right) \, .
\end{equation}

The smallness of the {absolute neutrino mass scale} may have different sources. In this paper we take the point of view of building low-energy renormalizable setups that are technically natural. In this respect, the lightness of $M_\Delta$ ({in the vicinity of the electroweak scale})
and the smallness of the {couplings} of the SM-singlet $\sigma$ with the doublet and triplet states (among which is $\lambda_6$) are a {required prerequisite} (see \sect{Naturalness}). The triplet Yukawa coupling $Y_\Delta$ is also constrained by tree-level LFV (see \sect{scalarph}).

\section{PQ extended Babu-Julio model}\label{sec:Babu-Julio}

We shall here introduce a simple PQ extension of the model of Ref.~\cite{Babu:2013pma}, 
which is a special case of the general Zee model \cite{Zee:1980ai}. 
For convenience, we display the field content and the relative PQ charges in \Table{fcBJ}, 
where $\alpha=1,2,3$ denotes the family index and $X_u + X_d \neq 0$ in order to obtain a non-vanishing QCD anomaly. The non-universal assignment of the PQ charges in the leptonic sector replaces the role of the $Z_4$ symmetry employed in \cite{Babu:2013pma}.

\begin{table}[thbp]
  \centering
  \begin{tabular}{@{} |c|c|c|c|c|c| @{}}
 \hline
    Field & Spin & $SU(3)_C$ & $SU(2)_L$ & $U(1)_Y$ & $U(1)_{PQ}$ \\ 
 \hline
    $q_L^\alpha$ & $\frac{1}{2}$ & 3  & 2 & $+\frac{1}{6}$ & 0\\ 
    $u_R^\alpha$ & $\frac{1}{2}$ &  3  & 1 & $+\frac{2}{3}$ & $X_u$ \\ 
    $d_R^\alpha$ & $\frac{1}{2}$ &  3  & 1 & $-\frac{1}{3}$ & $X_d$ \\ 
    $\ell^1_L$ & $\frac{1}{2}$ &  1  & 2 & $-\frac{1}{2}$ & $X_{\ell_1}$ \\ 
    $\ell^{2,3}_L$ & $\frac{1}{2}$ &  1  & 2 & $-\frac{1}{2}$ & $X_{\ell_{2,3}}$ \\ 
    $e_R^\alpha$ & $\frac{1}{2}$ &  1  & 1 & $-1$ & $X_e$ \\
    $H_u$ & 0 & 1 & 2 & $-\frac{1}{2}$ & $-X_u$ \\
    $H_d$ & 0 &  1  & 2 & $+\frac{1}{2}$ & $-X_d$ \\
    $h^+$ & 0 &  1  & 1 & +1 & $X_h$ \\
    $\sigma$ & 0 &  1  & 1 & 0 & $X_\sigma$ \\
 \hline
  \end{tabular}
  \caption{\label{fcBJ} Field content and charge assignment of the PQ extended Babu-Julio model.}
\end{table}

\subsection{Lagrangian}

The relevant Yukawa Lagrangian {reads} \cite{Babu:2013pma}
\bea
\label{YukawaBabuJulio}
- \mathcal{L}^{\rm{BJ}}_{\rm{Y}} &=& 
Y_u \, \overline{q}_{L} u_{R} H_u 
+ Y_d \, \overline{q}_{L} d_{R} H_d 
+ Y_{\beta\alpha} \, \overline{\ell}_{L}^\beta e_{R}^\alpha H_d    \nn\\
&&+ Y_{1\alpha} \, \overline{\ell}_{L}^1 e_{R}^\alpha \tilde{H}_u 
+ f_{23} (\ell_L^{2})^T  i \tau_2 C \ell^3_L h^+
+ \rm{h.c.} \, , 
\eea
where $\alpha=1,2,3$ and $\beta = 2,3$ label interaction eigenstates. The analogous contractions in the quark sector are understood. 

In the scalar potential, the terms  
$\sigma^2 \tilde{H}^\dag_u H_d$
and
$\sigma h^- H_u^\dag H_d$ are required. The {former} is needed
 to assign a non-vanishing PQ charge to the singlet $\sigma$, while the {latter is needed} to generate neutrino masses at one loop. 
 Notice that the simultaneous presence of 
these scalar interactions, together with $f_{23}$, breaks the lepton number {explicitly}.

The remaining part of the scalar potential contains only moduli {terms} and coincides with that of Ref.~\cite{Bertolini:1990vz}, namely:
\begin{align}
\label{BJ-scalarpot}
V_{\rm{BJ}} &= -\mu_1^2 \abs{H_u}^2 + \lambda_1 \abs{H_u}^4 -\mu_2^2 \abs{H_d}^2 + \lambda_2 \abs{H_d}^4  \\
&+ \lambda_{12} \abs{H_u}^2 \abs{H_d}^2 + \lambda_{4} \abs{H_u^\dag H_d}^2 \nonumber \\
& - \mu_3^2 \abs{\sigma}^2 + \lambda_3  \abs{\sigma}^4 
+ \lambda_{13} \abs{\sigma}^2 \abs{H_u}^2
+ \lambda_{23} \abs{\sigma}^2 \abs{H_d}^2 \nn \\
& + \abs{h}^2 \left( \mu_4^2 + \lambda_{41} \abs{H_u}^2 + \lambda_{42} \abs{H_d}^2 + \lambda_{43} \abs{\sigma}^2 \right. \nonumber \\
& \left. + \lambda_{44} \abs{h}^2 \right) 
+ \left( \lambda_5 \sigma^2 \tilde{H}^\dag_u H_d + \lambda_6 \sigma h^- H_u^\dag H_d + \text{h.c.} \right)  \, .  \nonumber
\end{align}
The couplings $\lambda_5$ and $\lambda_6$ can be set real by two independent rephasings of the fields. 
For a detailed discussion of the symmetry breaking patterns and the scalar spectrum we refer the reader directly to 
Sect.~3 of \cite{Bertolini:1990vz}. 

The Higgs content of the original model is extended by just the SM singlet $\sigma$. 
{The discussion here is rather similar to that in \sect{scalarspectrumTII} with the only difference that, given} 
the lepton family dependence of the PQ symmetry,
the axion {here} exhibits flavor non-diagonal couplings to leptons; {for more details see \sect{discussion}.} 

\subsection{PQ charges}

The {devised} $U(1)_{PQ}$ invariance of the Lagrangian leads to the following constraints on the PQ charges: 
\begin{align}
\label{PQeq1BJ}
- X_{\ell_{2,3}} + X_e - X_d &= 0 \, , \\
\label{PQeq2BJ}
- X_{\ell_{1}} + X_e + X_u &= 0 \, , \\
\label{PQeq3BJ}
2 X_{\ell_{2,3}} + X_h &= 0 \, , \\
\label{PQeq4BJ}
2 X_\sigma - X_u  - X_d &= 0 \, , \\
\label{PQeq5BJ}
X_\sigma - X_h + X_u  - X_d &= 0 \, .
\end{align}
{Solving these} in terms of $X_u$ and $X_d$ one obtains: 
\begin{align}
\label{PQchargesXuXdBJ}
&X_{\ell_{1}} = \frac{X_u}{4} + \frac{5 X_d}{4} \, , \quad
X_{\ell_{2,3}} = - \frac{3 X_u}{4} + \frac{X_d}{4} \, ,  \nn\\ 
&X_e = - \frac{3 X_u}{4}  + \frac{5 X_d}{4} \, ,  \
X_h = \frac{3 X_u}{2}  - \frac{X_d}{2} \, , \nn\\
&X_\sigma = \frac{X_u}{2}  + \frac{X_d}{2} \, . 
\end{align} 
{As before, cf.} \sect{PQchargesTII}, we 
require the orthogonality of the hypercharge and axion currents 
and fix the overall normalization of the charges by the condition
$X_\sigma= 1$,  
which yields
\begin{align}
\label{PQchargesxBJ}
&X_{\ell_1} = \frac{5 x^2 + 1}{2 (x^2 + 1)} \, , \ 
X_{\ell_{23}} = \frac{x^2 - 3}{2 (x^2 + 1)} \, , \nn\\
&X_e = \frac{5 x^2 - 3}{2 (x^2 + 1)}  \, , \quad
X_h = \frac{3 - x^2}{x^2 + 1}  \, . 
\end{align} 
The $X_u$ and $X_d$ charges are identical to those of the {TII model and, as such, they} are given in \eq{PQchargesx}.

\subsection{Neutrino masses}

\begin{figure}
\begin{center}
\includegraphics[width=0.9\columnwidth]{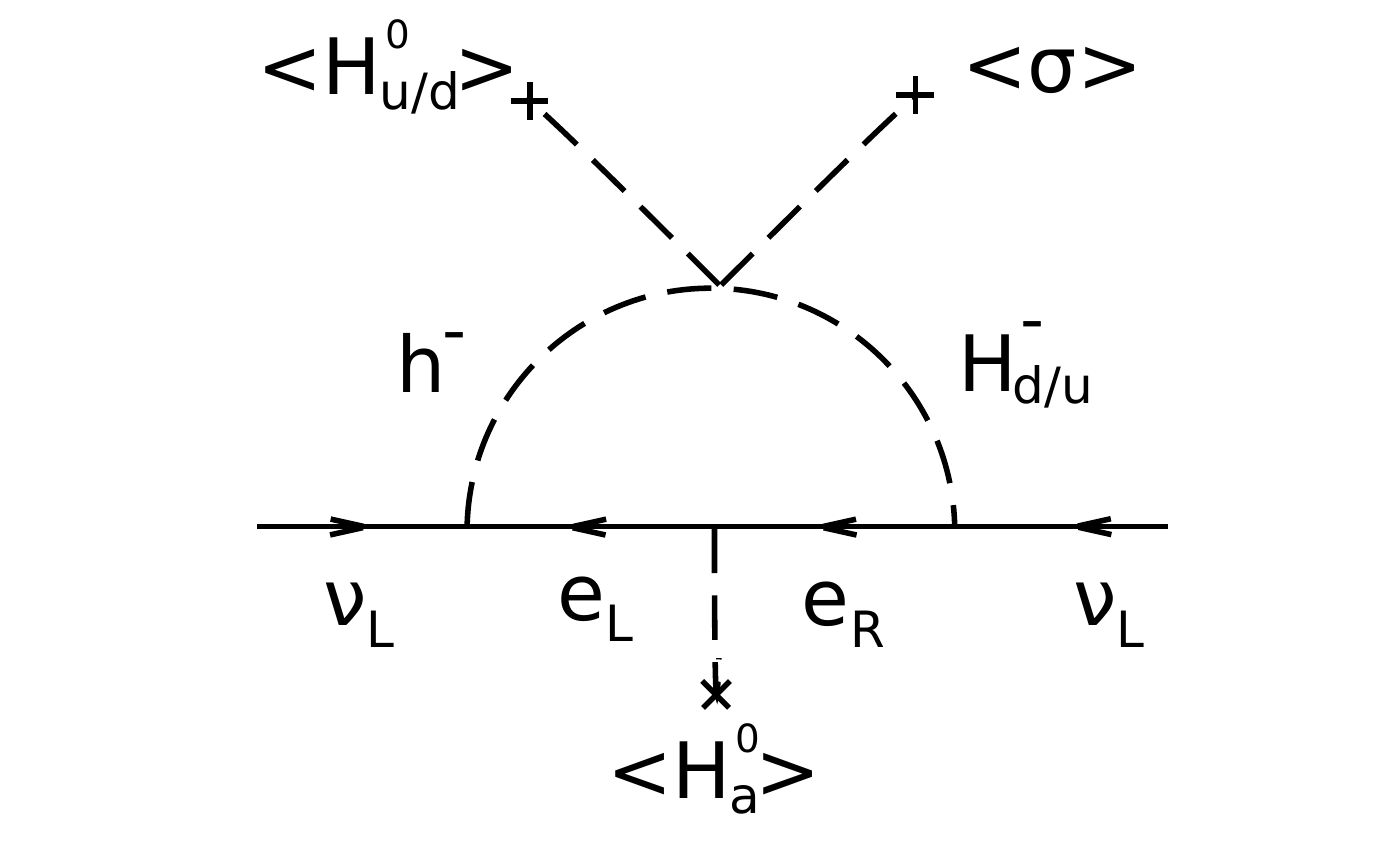}
\caption{Sample one-loop diagram responsible for the {Majorana} neutrino mass in the PQ extended Babu-Julio model.
Due to the family dependence of the PQ charges both Higgs doublets couple to the first leptonic family.}
\label{Fig:axbabujulio}
\end{center}
\end{figure}

The radiatively induced neutrino mass matrix is found to be \cite{Babu:2013pma} 
\begin{equation}\label{MnuBJ}
M^{\rm{BJ}}_\nu = \kappa \left( \hat{f} M_\ell^{\rm{diag}} \hat{Y}^T + \hat{Y} M_\ell^{\rm{diag}} \hat{f}^T \right) \, , 
\end{equation}
where 
$M_\ell^{\rm{diag}}$ is the diagonal charged-lepton mass matrix 
and $\hat{f}$ and $\hat{Y}$ are the Yukawa coupling matrices 
{transformed into} the mass basis {of the fields} running 
in the loop
(cf.~\fig{Fig:axbabujulio}). 
The main difference with respect to the $Z_2$-assisted Zee model \cite{Zee:1980ai,Wolfenstein:1980sy}, 
is that $\hat{Y}$ is non-diagonal, so that the antisymmetric texture of $\hat{f}$ is not transmitted 
into the neutrino mass matrix. 
This allows for non-vanishing diagonal entries at the leading order, a key feature for the consistency of the BJ model with the neutrino oscillation data.   
The overall factor $\kappa$ involves the loop function and 
is given by \cite{Babu:2013pma} 
\begin{equation}
\kappa = \frac{\sin 2 \gamma}{16 \pi^2} \log \left( \frac{M_1^2}{M_2^2} \right) \, , 
\end{equation}
{where $M_{1,2}$ are the masses of the physical charged scalar states and $\gamma$ denotes the mixing angle between $h^-$ and $H^-$ obeying (cf. \cite{Bertolini:1990vz})}
\begin{equation}
\sin 2 \gamma = \frac{2 \lambda_6 V_{\sigma} \sqrt{v_u^2 + v_d^2}}{M_1^2-M_2^2} \, .
\end{equation}
Interestingly, the structure of the neutrino mass matrix in \eq{MnuBJ} is very constrained. Albeit with non-vanishing diagonal entries the mass matrix turns out to be traceless and real so that
all the neutrino oscillation data can be described in terms of four real parameters~\cite{Babu:2013pma}. This leads to several predictions: 
the neutrino mass hierarchy is predicted to be inverted, the {Dirac} CP-violating phase {is fixed to} $\delta_{CP} = \pi$ {and, moreover, there is} a relation among the 
three mixing angles, namely $|U_{\tau 1}| =|U_{\tau 2}|$, {allowing one of them to be expressed} in terms of the other two. 
{The} consequences for neutrinoless double beta decay and LFV processes {have been} systematically worked out in Ref. \cite{Babu:2013pma}.

As in the TII case, the smallness of neutrino masses can have different sources. 
If the charged scalar states are not far from the electroweak scale, 
as suggested by {the} naturalness arguments,
the suppression must come from the scalar potential coupling $\lambda_6$ 
and/or the Yukawa matrices $\hat{Y}$ and $\hat{f}$. 
{Remarkably,} the smallness of the coupling $\lambda_6$ is {a necessary condition} for a technically natural spectrum (see \sect{Naturalness}), 
while the Yukawa couplings $\hat{Y}$ and $\hat{f}$ are sharply constrained by 
neutrino oscillation data and LFV processes \cite{Babu:2013pma}.

\section{PQ extended Zee-Babu model}\label{sec:Zee-Babu}

The last case we are going to consider is the PQ extension of the ZB model. 
The field content and the PQ charges are collected for convenience in \Table{fcZB}.

\begin{table}[thbp]
  \centering
  \begin{tabular}{@{} |c|c|c|c|c|c| @{}}
 \hline
    Field & Spin & $SU(3)_C$ & $SU(2)_L$ & $U(1)_Y$ & $U(1)_{PQ}$ \\
 \hline
    $q_L$ & $\frac{1}{2}$ & 3  & 2 & $+\frac{1}{6}$ & 0\\
    $u_R$ & $\frac{1}{2}$ &  3  & 1 & $+\frac{2}{3}$ & $X_u$ \\
    $d_R$ & $\frac{1}{2}$ &  3  & 1 & $-\frac{1}{3}$ & $X_d$ \\
    $\ell_L$ & $\frac{1}{2}$ &  1  & 2 & $-\frac{1}{2}$ & $X_\ell$ \\
    $e_R$ & $\frac{1}{2}$ &  1  & 1 & $-1$ & $X_e$ \\
    $H_u$ & 0 & 1 & 2 & $-\frac{1}{2}$ & $-X_u$ \\
    $H_d$ & 0 &  1  & 2 & $+\frac{1}{2}$ & $-X_d$ \\
    $h^+$ & 0 & 1  & 1 & +1 & $X_h$ \\
    $k^{++}$ & 0 &  1  & 1 & +2 & $X_k$ \\
    $\sigma$ & 0 &  1  & 1 & 0 & $X_\sigma$ \\
 \hline
  \end{tabular}
  \caption{\label{fcZB} Field content and charge assignment of the PQ extended Zee-Babu model. }
\end{table}

\subsection{Yukawa interactions}
The Yukawa Lagrangian of the ZB model reads \cite{Herrero-Garcia:2014hfa}
\begin{multline}
\label{ZeeBabuYukawa}
- \mathcal{L}^{\rm{ZB}}_{\rm{Y}} =
Y_u \, \overline{q}_{L} u_{R} H_u
+ Y_d \, \overline{q}_{L} d_{R} H_d
+ Y_e \, \overline{\ell}_{L} e_R H_d  \\
+ f  \, \ell^T_{L} C i\tau_2 \ell_{L} h^+
+ g \, e_R^T C e_R k^{++}
+ \rm{h.c.} \, ,
\end{multline}
where $Y_u$, $Y_d$, $Y_e$, $f$, and $g$ are matrices in {the} generation space (flavor indexes are understood). In particular,  $f$ is antisymmetric while $g$ is symmetric. 
\eq{ZeeBabuYukawa} yields the following relations among the PQ charges of the fields involved:
\begin{align}
\nonumber X_e-X_\ell - X_d&=0 \, , \\
\label{PQYuk}2X_\ell + X_h &=0 \, , \\
\nonumber 2X_e + X_k &=0 \, .
\end{align}

\subsection{Scalar potential}\label{SecScalPot}

{As before,} the scalar potential is restricted by the requirement of {the $U(1)_{PQ}$ invariance.} However, 
different assignments of the PQ charges allow for different terms. In particular, we {consider}
\begin{align}
V_{\rm{ZB}} &= -\mu_1^2|H_u|^2 + \lambda_1|H_u|^4 -\mu_2^2 |H_d|^2+ \lambda_2 |H_d|^4 \nn\\
&+ \lambda_{12} |H_u|^2|H_d|^2 + \lambda_4 |H_u^\dagger H_d|^2 \nn\\
&- \mu_3|\sigma|^2 + \lambda_3|\sigma|^4 + \lambda_{13}|\sigma|^2|H_u|^2+\lambda_{23}|\sigma|^2 |H_d|^2 \nn\\
&+ |h|^2 \left( \mu_4^2 + \lambda_{41} |H_u|^2 + \lambda_{42}|H_d|^2 + \lambda_{43} |\sigma|^2 +\lambda_{44} |h|^2 \right)\nn\\
&+ |k|^2\left( \mu_5^2 + \lambda_{51} |H_u|^2 + \lambda_{52}|H_d|^2 + \lambda_{53} |\sigma|^2\right. \nn\\
&+\left.\lambda_{54} |h|^2 +\lambda_{55} |k^2|\right)
+(V_{\mathrm{var}} + \mathrm{h.c.}) \, , 
\end{align}
{with two different forms of the $V_{\mathrm{var}}$ part therein}:
\begin{itemize}
\item[$(i)$] 
$V_{\mathrm{var}} = \lambda_5 \sigma^2 H_d^\dagger \tilde{H_u} + \lambda_6\sigma h^- H_u^\dagger H_d + \lambda_7 \sigma (h^+)^2 k^{--}$

In this case the additional equations restricting the PQ charges read
\begin{align}
\nonumber 2X_\sigma + X_u + X_d &= 0 \, ,\\
\label{PQV1}X_\sigma - X_h + X_u - X_d &= 0 \, ,\\
\nonumber X_\sigma + 2 X_h - X_k &= 0 \, .
\end{align}
These, together with \eq{PQYuk}, represent a system of six equations 
for six variables (letting, e.g.,~$X_d$ to be a free parameter): $X_\sigma$, $X_h$, $X_k$, $X_e$, $X_\ell$ and $X_u$. 
Were these equations linearly independent, {there} would be a unique solution (up to the normalization of $X_d$) 
proportional to the SM hypercharge and, hence, $X_\sigma=0$. 
Consequently, in order {for $\sigma$} to get a non-trivial PQ charge, the system of equations {(\ref{PQYuk}) and (\ref{PQV1}) must} be linearly dependent, 
which is indeed the case.
Hence, we are allowed to set $X_\sigma$ non vanishing and we get
\begin{align}
\nonumber &X_h=-2X_d-X_\sigma \, , \quad X_k=-4X_d-X_\sigma \, , \\ 
\label{SolPQ1} &X_e=2X_d + \frac{1}{2}X_\sigma \, , \quad X_\ell=X_d+ \frac{1}{2}X_\sigma \, , \\
\nonumber & X_u = -X_d - 2X_\sigma \, .
\end{align}

Notice that the shape of $V_{\mathrm{var}}$ in case $(i)$ is, in a sense exceptional as it yields a linearly dependent set of equations for the PQ charges. {Were we to replace $\sigma\to\sigma^\ast$ in any one (or any two)} of the terms in $V_{\rm var}$ above (which is allowed by SM symmetries) the resulting system of equations would {be linearly independent} leaving as the only solution the one with the PQ charges proportional to the SM hypercharge. In such a case, in order to obtain {a non trivial assignment of the PQ charges}, the number of conditions must be reduced by setting one of the couplings $\lambda_{5,6,7}$ to zero.

{Unlike} the term proportional to $\lambda_7$ {which is required by the two-loop ZB diagram (see Fig.~\eqref{Fig:axbabuzee}) the} term proportional to $\lambda_6$ is, strictly speaking, not necessary (as a matter of fact, it induces an additional one-loop contribution to 
neutrino masses as in the Zee model~\footnote{Notice that this term is not present in the original ZB model, 
which features only one Higgs doublet \cite{Zee:1985id,Babu:1988ki}.}).
Hence, in the following we will focus on an alternative PQ-charge assignment where $\lambda_6=0$.

\item[$(ii)$] $V_{\mathrm{var}}  = \lambda_5 (\sigma^\ast)^2 H_d^\dagger \tilde{H_u} + \lambda_7 \sigma (h^+)^2 k^{--} $

In this case the equations relating the PQ charges read
\begin{align}
\label{PQV2} -2X_\sigma + X_u + X_d &= 0 \, , \nonumber \\
X_\sigma + 2 X_h - X_k &= 0 \, .
\end{align}
By solving Eqs.~(\ref{PQYuk}) and (\ref{PQV2}) in terms of $X_u$ and $X_d$ we get:
\begin{align}
&X_{\ell} = \frac{X_u}{4} + \frac{5 X_d}{4} \, , \quad X_e = \frac{X_u}{4}  + \frac{9 X_d}{4} \, , \nn \\
&X_h = -\frac{X_u}{2}  - \frac{5X_d}{2} \, , \quad X_k = -\frac{X_u}{2}  - \frac{9 X_d}{2} \, , \nn \\
& X_\sigma = \frac{X_u}{2}  + \frac{X_d}{2} \, .  
\end{align} 
If we again require the orthogonality of the hypercharge and axion currents 
and choose the normalization $X_\sigma = 1$, we obtain the following expressions in terms of $x=v_u/v_d$
\begin{align}
&X_\ell = \frac{5 x^2 + 1}{2 (x^2 + 1)}\, , \quad X_e = \frac{9 x^2 +1}{2 (x^2 + 1)} \, , \nn \\
&X_h = \frac{- 5 x^2 -1}{x^2 + 1}\,, \quad X_k = \frac{- 9 x^2 -1}{x^2 + 1}\, ,
\end{align}
with $X_u$ and $X_d$ given as in \eq{PQchargesx}.

\end{itemize}
From now on we will only consider the shape of the scalar potential in case $(ii)$ {(that leads to the extended ZB model)}, 
and we will briefly comment on the differences with respect to {the case $(i)$ in  \sect{SecNeutrinoMassesZB}.}

\subsection{Scalar spectrum\label{sec:ZBspectrum}}

{Adopting the notation of Eqs.~\eqref{expHu}, \eqref{expHd} and \eqref{expsigma} for the relevant fields, the stationarity} conditions read
\begin{align}
\! - \lambda_5 v_d V_\sigma^2 + v_u\left(\lambda_{12} v_d^2 + \lambda_{13} V_\sigma^2 + 2\lambda_1 v_u^2 -\mu_1^2\right)& = 0 \, ,\\
\! - \lambda_5 v_u V_\sigma^2 + v_d\left(\lambda_{12} v_u^2 + \lambda_{23} V_\sigma^2 + 2\lambda_2 v_d^2  -\mu_2^2\right)& = 0 \, ,\\
\! V_\sigma\left(\lambda_{13} v_u^2 + \lambda_{23} v_d^2 + 2 \lambda_3 V_\sigma^2 - 2 \lambda_5 v_d v_u - \mu_3^2\right)&=0 \, .
\end{align}
{Using these and expanding around the vacuum configuration} 
the mass matrix of the neutral scalar fields in the $\{h_u^0, h_d^0, \sigma^0\}$ basis turns out to be
\begin{multline}
M_{\rm{S}}^2 = V_\sigma^2 \\
\times \left(\begin{array}{ccc}
\lambda_5 \frac{v_d}{v_u} + 4\lambda_1 \frac{v_u^2}{V_\sigma^2} &  - \lambda_5 + 2\lambda_{12}\frac{v_d v_u}{V_\sigma^2}& 2 \frac{\lambda_{13} v_u- \lambda_5 v_d }{V_\sigma}\\
- \lambda_5 + 2\lambda_{12}\frac{v_d v_u}{V_\sigma^2}  & \lambda_5\frac{ v_u}{v_d} + 4\lambda_2 \frac{ v_d^2}{V_\sigma^2} & 2 \frac{\lambda_{23} v_d- \lambda_5 v_u}{V_\sigma}\\
 2 \frac{\lambda_{13} v_u- \lambda_5 v_d }{V_\sigma}&2 \frac{\lambda_{23} v_d- \lambda_5 v_u}{V_\sigma}& 4\lambda_3
\end{array}\right).
\end{multline}
Assuming $V_\sigma \gg v \equiv \sqrt{v_u^2 + v_d^2}$ and keeping only {$\mathcal{O}(V_\sigma^2)$} terms, the set of eigenvalues 
of $M_{\rm{S}}^2$ reads $\{0, \frac{v^2}{v_d v_u}\lambda_5 V_\sigma^2, 4\lambda_3 V_\sigma^2\}$. The {$\mathcal{O}(v V_\sigma)$ perturbations do not affect this result at the first order and, hence,} we conclude that the eigenvalues of $M_{\rm{S}}^2$ are
\beq
\label{eigenM2Sexp}
\left\{ \mathcal{O}(v^2), \, \frac{v^2}{v_d v_u}\lambda_5 V_\sigma^2+ \mathcal{O}(v^2),\, 4\lambda_3 V_\sigma^2 + \mathcal{O}(v^2)\right\}.
\eeq
 
Next, the pseudoscalar mass matrix has two eigenvalues equal to zero: the combination $-\frac{v_u}{v} \eta_u^0 + \frac{v_d}{v} \eta_d^0$ corresponds to the goldstone boson (GB) ``eaten'' by the $Z$, whereas $\frac{v_u}{V_\sigma}\frac{v_u v_d}{v^2} \eta_u^0 + \frac{v_d}{V_\sigma}\frac{v_u v_d}{v^2} \eta_d^0  + \eta_\sigma^0 $ is the axion -- the GB corresponding to the breaking of the PQ symmetry. The third pseudoscalar  eigenstate $-\frac{v_d}{v} \eta_u^0 - \frac{v_u}{v} \eta_d^0 + \frac{2v_uv_d}{v V_\sigma}\eta_\sigma^0$ acquires mass
\begin{equation}
m^2_{\rm{PS}} = \lambda_5\left(4 v_u v_d + \frac{v^2}{v_u v_d}V_\sigma^2\right) \, .
\end{equation}
Among the singly charged scalars, the GB ``eaten'' by the $W$ corresponds to the combination $\frac{1}{v}(-v_u h_u^+ + v_d h_d^+)$, whereas the orthogonal combination $\frac{1}{v}(v_d h_u^+ + v_u h_d^+)$ acquires mass
\begin{equation}
m^2_{H^+}=\lambda_4 v^2 + \lambda_5 \frac{v^2}{v_u v_d} V_\sigma^2 \, .
\end{equation}
For $\lambda_6 = 0$ the singlet scalar $h^+$ does not mix with $H^+$ 
and its mass reads
\begin{equation}
m^2_{h^+}=\mu_4^2 + \lambda_{41}v_u^2 + \lambda_{42}v_d^2 + \lambda_{43} V_\sigma^2 \, .
\end{equation}
Finally, the {mass of the} doubly-charged scalar $k^{++}$ {is given by}
\begin{equation}
m^2_{k^{++}}=\mu_5^2 + \frac{1}{2}\left(\lambda_{51}v_u^2 + \lambda_{52}v_d^2\right) + \lambda_{53} V_\sigma^2 \, . 
\end{equation}
Let us note that if $\lambda_{i3}, \lambda_5 \lesssim
\mathcal{O}(\frac{v^2}{V_\sigma^2})$ (with $i$ running over all the states but the SM singlet {$\sigma$}), 
all the scalars {besides one (the $\sigma$-dominated state with eigenvalue proportional to $\lambda_3$ in \eq{eigenM2Sexp})}, 
have tree-level masses of the order of the electroweak scale. 
As a matter of fact when $\lambda_{i3}, \lambda_5, \lambda_7 \ll 1$ the singlet
$\sigma$ is only weakly coupled to the other fields. 
As we shall discuss in \sect{Naturalness} this is a technically natural limit 
associated with the emergence of an extra Poincar\'e symmetry in the action. 
Notice that even though $\lambda_7$ does not enter the spectrum at {the} tree level, 
it is expected to contribute at one loop {(for instance,~to the mass of $k^{++}$).}
This {effect} is estimated to be roughly $\frac{1}{16\pi^2}\lambda_7^2
V_\sigma^2$ and hence, {the requirement} that the mass of $k^{++}$ is around the electroweak scale corresponds to 
$\lambda_7\lesssim 4 \pi \times \mathcal{O}(\frac{v}{V_\sigma})$ {which is consistent} with the decoupling of the heavy singlet.

\subsection{Neutrino Masses}
\label{SecNeutrinoMassesZB}
\begin{figure}
\begin{center}
\includegraphics[width=.9\columnwidth]{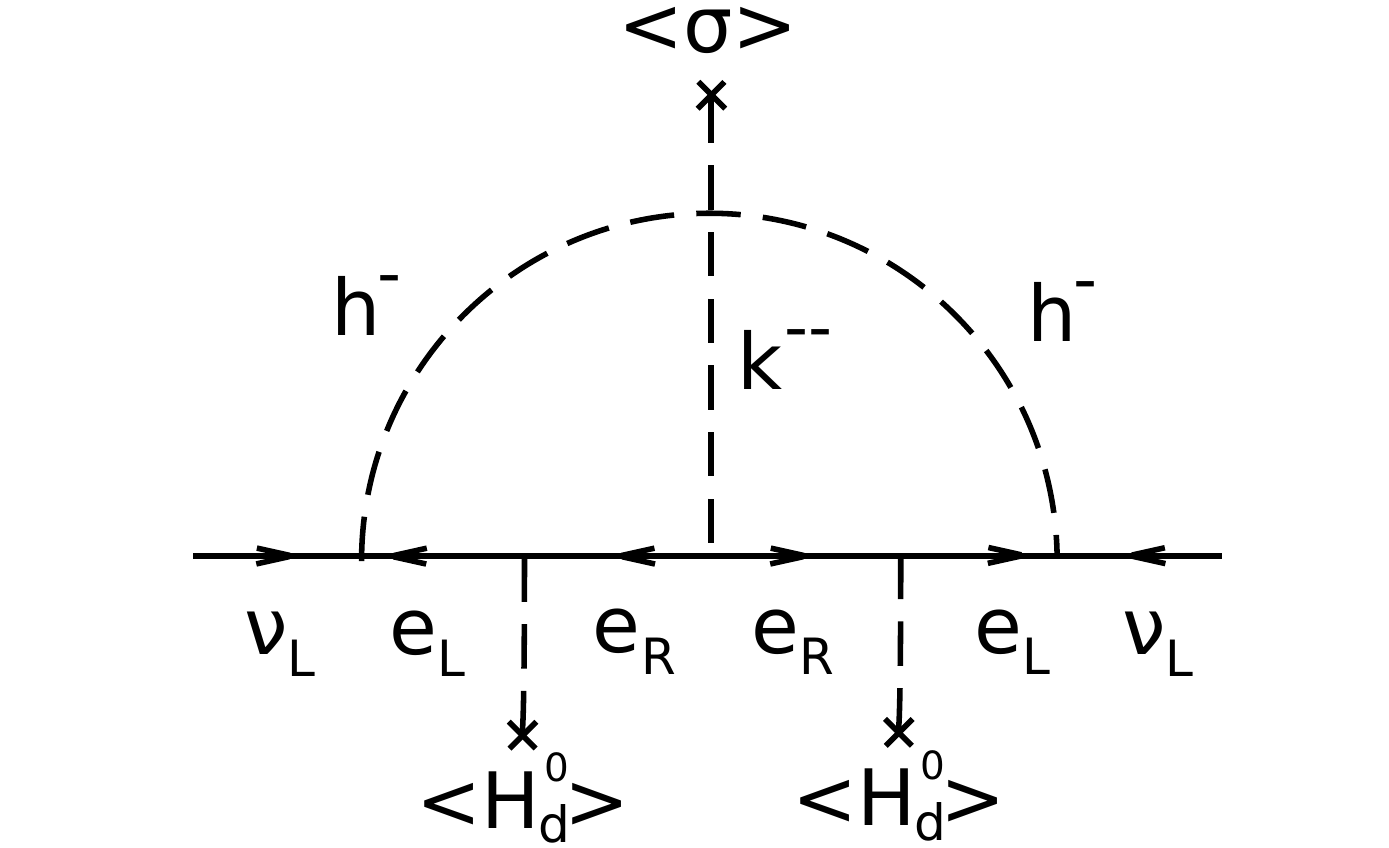}
\caption{Two-loop diagram responsible for the neutrino {Majorana} mass in the PQ extended Zee-Babu model.}
\label{Fig:axbabuzee}
\end{center}
\end{figure}

Focusing on the case {with} $\lambda_6 = 0$, 
the only radiative contribution to {the} neutrino masses is a 
two-loop diagram \`a la ZB (see Fig.~\ref{Fig:axbabuzee}), 
which yields \cite{Herrero-Garcia:2014hfa}
\begin{equation}\label{MnuZeeBabu}
(M_{\nu}^{\rm{ZB}})_{ij} = 16 \lambda_7 V_\sigma f_{ia} m_a g^{\ast}_{ab}I_{ab}m_b f_{jb} \, , 
\end{equation}
with $m_a$ denoting the {$a$-th charged lepton mass.} For $m_a \ll m_{h^+},m_{k^{++}}$, the loop function reads
\begin{equation}
\label{IntegralZeeBabu}
I_{ab} \simeq I = \frac{1}{(16\pi^2)^2}\frac{1}{M^2}\frac{\pi^2}{3} \tilde{I}(r) \, , 
\end{equation}
with $M\equiv \max( m_{h^+},m_{k^{++}})$ and 
\beq
\tilde{I}(r) = \Bigl\{ \begin{array}{ll}
1+\frac{3}{\pi^2}(\log^2 r -1)& r\gg 1\\
 1& r\to 0 
\end{array} \, , 
\eeq
where $r\equiv m_{k^{++}}^2 / m_{h^+}^2$. For the exact analytic result see~\cite{McDonald:2003zj}.

An important feature of the ZB model is that the lightest neutrino is predicted to be massless.
Indeed, since $f$ is antisymmetric, $\det f = 0$ (for three generations) and, hence, $\det M^{\rm{ZB}}_\nu = 0$. 

As in the previous cases (TII and BJ), the smallness of neutrino masses can be due to different factors. 
Taking into account the strong bounds on the Yukawa couplings $f$ and $g$ {coming from the} 
LFV processes (see \sect{scalarph}), it turns out that the assumption 
$\lambda_7\lesssim 4 \pi \times \mathcal{O}(\frac{v}{V_\sigma})$, {tailored} to keep the non-singlet scalars 
at the electroweak scale, ensures also the {correct absolute neutrino mass scale} \cite{Herrero-Garcia:2014hfa}.

Finally, we briefly comment on the case $\lambda_6\neq 0$. In such a {setting} there is an extra one-loop 
contribution to {the} neutrino masses, similar to the one in Fig.~\ref{Fig:axbabujulio} ({the relevant} expression can be 
found in Eqs.~(25)-(26) of \cite{Bertolini:1990vz}). 
As already mentioned, the original Zee model is excluded by neutrino data and,
in order to obtain a viable neutrino texture, the size of such a one-loop diagram must be 
comparable with the two-loop expression in \eq{MnuZeeBabu}, thus introducing a fine-tuning in the couplings. Let us also note that 
$\lambda_6\neq 0$ introduces a tree-level mixing between the {charged $SU(2)_{L}$-doublet and singlet} scalars that affects \eq{IntegralZeeBabu}. 
At variance with the ZB model, the lightest neutrino is no longer massless. {In this study, we will not pursue the analysis of this hybrid model any further}.

\section{Discussion}
\label{discussion}

The three setups presented in the previous sections share a {number} of common features which we shall {briefly summarize} here. 
{In particular,} all three models {contain} a DFSZ invisible axion with a tiny coupling to neutrinos~\cite{Bertolini:1990vz,Arason:1990sg}. It is noticeable that, at variance with the TII and ZB extended models, {the axion in the BJ case} exhibits flavour violating couplings to the leptons 
of the same order of the {flavour-diagonal} ones:
\begin{multline}
\mathcal{L}_{a\ell\ell} = 
- X_{\ell_{2,3}} \frac{\partial_\mu a}{f_a} \left[ ( \overline{e}^i_L  \gamma^\mu e^i_L ) + ( \overline{\nu}^i_L  \gamma^\mu \nu^i_L ) \right] \\
- X_{e} \frac{\partial_\mu a}{f_a} \left[ ( \overline{e}^i_R  \gamma^\mu e^i_R ) \right] \\
- (X_{\ell_1} - X_{\ell_{2,3}}) \frac{\partial_\mu a}{f_a} \left[ ( \overline{e}^i_L  \gamma^\mu (U^{e\dag}_L)^{i1} (U^e_L)^{1j}  e^j_L ) \right] \\
+ \left. ( \overline{\nu}^i_L  \gamma^\mu (U^{\nu\dag}_L)^{i1} (U^\nu_L)^{1j}  \nu^j_L ) 
\right] \, , 
\end{multline} 
that, up to a total derivative, can be written as
\begin{multline}
\mathcal{L}_{a\ell\ell} = 
 i\frac{a}{f_a} \left[\left(X_e-X_{\ell_{2,3}}\right) m_i^e\ \overline{e}^i \gamma_5 e^i  -X_{\ell_{2,3}} m_i^\nu\ \overline{\nu}^i  \gamma_5 \nu^i \right]\\
- i\left(X_{\ell_1} - X_{\ell_{2,3}}\right) \frac{a}{f_a} \left[  (U^{e\dag}_L)^{i1} (U^e_L)^{1j} \  \overline{e}^i  \left(\frac{m_j^e-m_i^e}{2} \right. \right. \\
+  \left.  \frac{m_j^e+m_i^e}{2} \gamma_5  \right) e^j \\
+ \left.  (U^{\nu\dag}_L)^{i1} (U^\nu_L)^{1j} \ \overline{\nu}^i  \left(\frac{m_j^\nu-m_i^\nu}{2}  +  \frac{m_j^\nu+m_i^\nu}{2} \gamma_5  \right) \nu^j 
\right] \, .
\end{multline} 
where {$a$  denotes the axion field} and $f_a = \sqrt{2} V_\sigma$. The mass eigenstates $e^i_L$, $\nu^i_L$ ($i=1,2,3$) are connected to the interaction basis  $e^\alpha_L$, $\nu^\alpha_L$ ($\alpha=1,2,3$) 
via the relations $e^\alpha_L = (U^e_L)^{\alpha i} e^i_L$ and $\nu^\alpha_L = (U^\nu_L)^{\alpha i} \nu^i_L$. 
The equations of motion for Weyl fermions with a {Majorana} mass term are used and the axion neutrino couplings are written in terms of the {Majorana} mass eigenstates~\cite{Bertolini:1987kz}.  
Present laboratory and astrophysical limits on flavor violating interactions {do not seem to imply any} constraints on the PQ scale stronger than those obtained from the diagonal interactions~\cite{Jaeckel:2013uva}. On the other hand, the presence of lepton flavor 
{violating} interactions of the axion in the extended BJ model deserves further detailed scrutiny. 

The DFSZ invisible axion framework suffers {from} the domain wall problem (the non-perturbative instanton potential breaks {the $U(1)_{PQ}$} explicitly  to a $Z_{N_q}$ discrete symmetry {where $N_q$ is} the number of quark flavors). The standard cosmological solution is {then the assumption of} a low reheating temperature (see~\cite{Kim:1986ax} for a comprehensive discussion).

\subsection{Naturalness}
\label{Naturalness}

An interesting feature of {all} the models considered in this {study} is the fact that the hierarchy between 
the electroweak and the PQ scales can be made technically natural and stable against radiative corrections. 
Let us consider, for definiteness, the case of the PQ extended TII model. 
At {the} tree level, 
{the} hierarchy between the PQ and the electroweak scale
can be obtained without fine-tuning among the scalar potential parameters of \eq{TII-scalarpot} 
by requiring the ultraweak limit 
\begin{equation}
\label{lambdahierarchies}
\lambda_{i3}, \lambda_5 \sim \mathcal{O}\left(\frac{v^2}{V_\sigma^2}\right) 
\quad \text{and} \quad 
\lambda_6 \sim \mathcal{O}\left(\frac{v_\Delta}{V_\sigma}\right) \, ,
\end{equation}
where the last equation is set by the {stationarity condition (\ref{stateqvD})} and 
$i$ is running over all the scalar multiplets but the SM singlet 
(all non-singlet mass parameters are taken at the weak scale).
As a matter of fact, this guarantees that the heavy (PQ-scale) neutral singlet 
decouples from the rest of the spectrum {(see \app{scalarspectrumII} and  \sect{sec:ZBspectrum})}. 
It is noteworthy that 
the ultraweak limit $\lambda_{i3}, \lambda_5, \lambda_6 \ll 1$ 
is associated with the emergence of an 
additional Poincar\'e symmetry of the action \cite{Volkas:1988cm,Bertolini:1990vz} 
(see \cite{Foot:2013hna} for a recent discussion) 
which makes this limit perturbatively stable. It is readily verified that the renormalization of the 
couplings connecting the ``light'' and ``heavy'' sectors is as a set multiplicative (the relevant beta functions exhibit a fixed point for vanishing couplings). The hierarchy among the ultraweak couplings in \eq{lambdahierarchies} is stable since $\lambda_6^2 \ll \lambda_{i3}$. The couplings $\lambda_5$ and $\lambda_6$ are themselves multiplicatively renormalized since lepton number is restored when one of them is vanishing. 
The naturalness requirement, together with the constraints coming from {the} LFV phenomenology, allows us  
to reproduce in all three setups {above} the correct neutrino mass scale {together} with a {number} of new scalar states {in} the reach of present collider searches.

\begin{figure}[h]
\centering
\includegraphics[angle=0,width=8.cm]{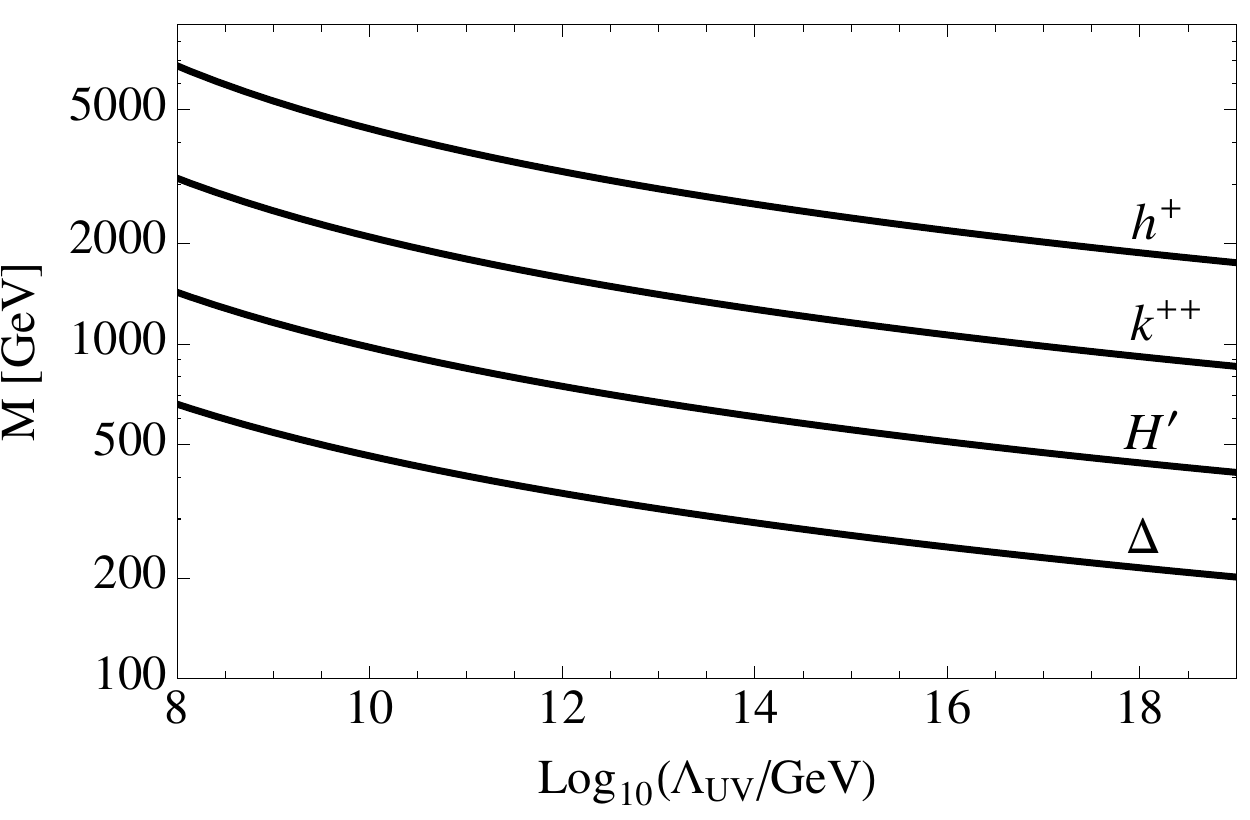}
\caption{\label{FNbounds} 
Naturalness bounds on the mass of the exotic scalar multiplets present in the different PQ extended neutrino mass models, as a function of the ultraviolet cut-off, {see \Eq{naturalnessformula}.}}
\end{figure}

Beyond the tree level, we can set naturalness bounds on the mass scale $M$ of the {non-SM} scalar multiplets 
by requiring that the finite two-loop gauge corrections to the Higgs mass parameter 
$m^2$ do not exceed the Higgs pole mass. 
{Adopting} the notation of Ref.~\cite{Farina:2013mla}, 
{each} complex scalar multiplet with quantum numbers 
$(n,Y)$ under $SU(2)_L \otimes U(1)_Y$ {contributes by} 
\begin{multline}
\label{naturalnessformula}
\delta m^2 (\bar\mu) = - \frac{n M^2}{(4 \pi)^4} 
\left( \frac{n^2-1}{4} g_2^4 + Y^2 g_Y^4 \right) \\ \times \left( \frac{3}{2} \ln^2 \frac{M^2}{\bar\mu^2} 
+ 2 \ln \frac{M^2}{\bar\mu^2} + \frac{7}{2} \right) \, ,  
\end{multline}
where $\bar\mu$ is the the renormalization scale in the $\overline{\rm{MS}}$ scheme, 
to be identified with the 
the cutoff of the effective theory $\Lambda_{\rm{UV}}$ (which in our setup is the onset of gravity, 
since we require the decoupling of the PQ-scale scalar singlet).
The naturalness bounds are shown in \fig{FNbounds}, where {we display the constraints on the individual contributions of  
the} scalar triplet $\Delta$, {the} extra Higgs doublet $H'$, 
{the} singly-charged scalar $h^+$ and {the} doubly-charged scalar $k^{++}$, respectively. 
For a fully natural spectrum, the new states are expected to live not too far from the 
electroweak scale, say below 5 TeV to 200 GeV depending on the multiplet considered 
and the value of $\Lambda_{\rm{UV}}$.

A few comments about gravity are in order. Even assuming for the time being no massive Planckian states
one should wonder whether the presence of the
heavy SM singlet might give rise to gravity-mediated radiative corrections that destabilize 
the light Higgs mass. In Refs.~\cite{Farina:2013mla,deGouvea:2014xba} it is pointed out that a finite gravity-mediated contribution arises
at the three-loop order and it is estimated as $\delta m^2 \approx \tfrac{Y_t^2 G_N^2 M^6}{(4 \pi)^6}$. 
This yields a naturalness bound on the singlet mass $M \lesssim 10^{14}$ GeV, 
which is {compatible} with the experimentally allowed range of the PQ scale.
On the other hand, {gravity-induced} corrections to the Higgs mass 
are generally expected to arise at the two-loop level due to the breaking of the Higgs shift symmetry by SM interactions 
(one-loop contributions in {the} minimally-coupled gravity are derivatively suppressed)
and, {on purely dimensional grounds,} they can be estimated to be $\delta m^2 \approx \tfrac{G_N \Lambda_G^4}{(4 \pi)^4}$. This raises
an issue of naturalness already at $\Lambda_G \approx 10^{11}$ GeV, thus questioning the stability of the
simple setups discussed here (as well {as that} of the SM). Softened gravity may be invoked, but still the appearance of Landau poles
at trans-Planckian energies remains in principle an issue~\cite{Giudice:2014tma}. Total asymptotic freedom of the low-energy setup is invoked as a solution~\cite{Giudice:2014tma,Holdom:2014hla}.  

Another potential issue related to gravity is the possibility of {an} 
explicit breaking of the global $U(1)_{PQ}$ due to Plank-scale physics. This {may induce a shift of} the vacuum of the axion potential 
{and, thus, endanger} the PQ solution of the strong CP problem~\cite{Kamionkowski:1992mf,Kallosh:1995hi,Holman:1992us}.  
The authors of \cite{Dvali:2013cpa} argue that the presence of {Majorana} neutrinos may provide a protection for the PQ mechanism, that 
leads to a connection between the upper bound on neutrino masses and the onset of gravitational effects at the scale $\Lambda_G$.
We plan to systematically address the potential issues related to gravity in a future work.

\subsection{Electroweak vacuum stability}
\label{VacuumStability}

{An added value} of the models considered in this {study}, is the fact that scalar extensions 
of the SM are {tailor-made} to improve on the electroweak vacuum stability. This issue has been discussed at length 
in the literature (see, {for instance},~\cite{Lebedev:2012zw,EliasMiro:2012ay}), and we just briefly recall the argument here.
The key effect is the contribution of the new scalars (through, e.g., {the Higgs portal couplings}) 
to the {running} of the Higgs quartic coupling $\lambda_H$ (see, {for example},~Refs.~\cite{Chao:2012mx,Dev:2013ff,Kobakhidze:2013pya} 
for analysis in the context of the Type-II seesaw). 
These corrections contribute positively to the {beta-function} of $\lambda_H$ and, hence, they tend to stabilize the Higgs potential 
if they enter the running below the instability scale.\footnote{The instability scale of the SM effective potential is a gauge dependent quantity \cite{DiLuzio:2014bua}. A gauge-invariant criterium to include the effects of new physics can be devised~\cite{Andreassen:2014gha}. 
} Another interesting possibility for the cases at hand (pointed out in~\cite{EliasMiro:2012ay}) is that {a} heavy SM singlet $\sigma$ may stabilize the Higgs potential through a large threshold effect, $\lambda_H \rightarrow \lambda_H - \lambda_{H\sigma}^2/\lambda_\sigma$. 
In our case, however, we advocate the ultraweak limit 
$\lambda_{H\sigma} \lesssim v^2/V_{\sigma}^2$ while leaving the $\sigma$ self-interactions unconstrained. 
We are therefore bound to the running effect alone which suffices in the extended setups here considered.
A detailed analysis {of these matters} is beyond the scope of the present paper and will be the subject of a future study.

Let us finally remark that even if the instability of the electroweak vacuum is, strictly speaking, not an issue as long as {the lifetime of the false vacuum is long} enough to comply with the age of the universe, the fate of the electroweak vacuum depends on the cosmological history and 
there are inflationary setups where the meta-stability of the {vacuum} may be an issue~\cite{Espinosa:2007qp}. 

\subsection{The extended scalar sector}
\label{scalarph}

The scalar sectors of the models here discussed are commonly extended to feature two Higgs doublets and a complex singlet (in addition to one $Y=1$  triplet in {the case of} TII, one singly charged singlet in {the BJ model} and two singly and doubly charged singlets in {the ZB setting}). With the exception of the light axion field and its scalar partner whose {masses are} driven by the PQ-breaking scale, all other physical scalars may live at the TeV scale, {yet be compatible with all the bounds from} the low-energy phenomenology and collider physics. 

Due to the hierarchy among the PQ, the electroweak and {the} triplet VEVs ({following from} the constrained form of the scalar potential), the weak scale neutral Higgs sector of the three models overlaps to a large extent with that of {the} two Higgs doublet extension of the SM
(see \app{scalarspectrumII}). Correspondingly, {we may obtain one of the physical neutral scalars to behave as the SM Higgs by invoking} the decoupling limit $m_A^2\gg |\lambda_i|v^2$~\cite{Gunion:2002zf} (where $A$ is the pseudo-scalar field and $\lambda_i$ are the relevant scalar couplings). On the other hand, a SM-like Higgs does not {necessarily} imply {the ``true''} decoupling of $A$ (and the other physical doublet states), since {the extra} scalar couplings can be small enough to allow for a full spectrum at the TeV scale while {retaining} the $\alpha\simeq \beta -\pi/2$ decoupling relation among the {relevant} diagonalization angles~\cite{Gunion:2002zf}. 

These considerations {apply} to any of the models discussed here. 
Since a detailed phenomenological study of the different scalar sectors is beyond the scope of the present work we shall briefly 
comment upon some {of the} relevant common features {focusing mainly on} the lepton flavor phenomenology and collider signatures.

\paragraph{Lepton flavor violation.}\label{ParLFV}

{Unlike for the BJ case, in the TII and ZB models} the tree-level flavor violation in neutral currents is forbidden by the presence of the PQ symmetry. 
On the other hand, {in all three scenarios}, the extended scalar sector allows for {the} tree-level flavor violation in {the} charged currents. In particular, the couplings of the charged scalars to leptons, {closely related to the structure of the neutrino mass matrix,} play a crucial role in the {LFV} phenomenology of the models.

In the case of the extended TII model {the} LFV arises from the Yukawa coupling $Y_{\Delta}$, which is {directly} proportional to the neutrino mass matrix (cf.~\eq{neutrinomasses}). As an example,
in order to avoid the stringent constraints from the non-observation of $\mu^-\to e^- e^+ e^-$, 
the relation $|(Y_\Delta)_{\mu e}|^2|(Y_\Delta)_{ee}|^2\left(\frac{200\,\GeV}{m_{\Delta^{++}}}\right)^4\lesssim 10^{-12}$ must be satisfied \cite{Akeroyd:2009nu}. 
{This, in turn,} considerably constrains the shape of the neutrino mass matrix {in the case of our interest, i.e., when} the triplet mass is in the LHC range. 
Taking into account these constraints, Ref.~\cite{Akeroyd:2009nu} predicts branching ratios for $\tau\to l^-_i l^+_i l^-_i$ and $\mu\to e\gamma$ accessible to the upcoming experiments with {implication for neutrino physics}. The present bounds on these processes are weaker than those coming from {the} $\mu^-\to e^- e^+ e^-$ decay; {for instance,} the $\tau$ decay measurements yield the constraint $|(Y_\Delta)_{\tau i}|^2|(Y_\Delta)_{jk}|^2\left(\frac{200\,\GeV}{m_{\Delta^{++}}}\right)^4\lesssim 10^{-7}$ \cite{Akeroyd:2009nu}. Notice, however, that these bounds are {in general} more relevant for {accessing} the overall neutrino mass scale than the bound from $\mu^-\to e^- e^+ e^-$ which can be satisfied by tuning just one of the $Y_\Delta$ entries.

Similarly, in the extended ZB model, one can obtain strong bounds on the Yukawa couplings $f$ and $g$ in \eq{ZeeBabuYukawa}
from {the} LFV decays (together with {the} charged-currents universality and the lepton anomalous magnetic moments). 
The strongest bound is {again} due to the non-observation of $\mu^-\to e^+ e^- e^-$ and {it} requires 
$|g_{e\mu}g_{ee}^\ast| \left(\tfrac{200\,\GeV}{m_{k^{++}}}\right)^2 \lesssim 10^{-6}$ \cite{Herrero-Garcia:2014hfa}.
{The bounds on the $f$ couplings from other LFV processes (such as $\mu\to e\gamma$) are somewhat weaker~\cite{Herrero-Garcia:2014hfa}.}
The tightest of them, induced by the non-observation of $\mu\to e\gamma$ decay, reads $|f_{e\tau}^\ast f_{\mu\tau}|^2 \left(\frac{200\,\GeV}{m_{h^{+}}}\right)^4 +16 |g_{ee}^\ast g_{e\mu} + g_{e\mu}^\ast g_{\mu\mu} +g_{e\tau}^\ast g_{\tau\mu}|^2\left(\tfrac{200\,\GeV}{m_{k^{++}}}\right)^4 \lesssim 3 \times 10^{-9}$. Because of the complicated shape of these constraints, the {correlations with the structure of the neutrino masses} matrix \eqref{MnuZeeBabu} is not straightforward. On the other hand, a detailed numerical analysis~\cite{Herrero-Garcia:2014hfa} shows that combining these constraints with the neutrino oscillation data allows {one} to predict the shape of the $g$ matrix and determine the branching ratios for the $k^{++}$ decay. For $m_{k^{++}}\leq 400\,\GeV$ and normal neutrino hierarchy {the $k^{++}$ scalar} decays into $ee$ or $\mu\mu$ only, whereas for the same mass range and inverse hierarchy the decay channel $\mu\tau$ is non-negligible and $k^{\pm\pm}\to h^\pm h^\pm$ opens. Consequently, the discovery of a doubly charged scalar would either {rule out} the ZB model or provide {a testable information about the neutrino mass pattern}.

The {analysis of} the extended BJ model must take into account that both Higgs doublets couple to leptons 
and, hence, {the} tree-level LFV processes can be mediated by {the} neutral scalars. 
On the other hand, this model is highly predictive and it is sharply constrained by {the} neutrino data. Remarkably, a thorough numerical {analysis~\cite{Babu:2013pma}} shows that only two specific solutions for the lepton coupling matrix $\hat{Y}$ are possible. 
We refer the reader to the original paper for a detailed discussion. 

\paragraph{Collider physics}
The direct production searches {at} LEP set a lower bound on the mass of {the} singly charged scalars ($\Delta^+$ or $h^{+}$ in our models) 
of about $90$~GeV. For the case of the charged physical scalar of {the} two Higgs doublet extension {of the SM the} LHC does not add any new stringent constraint 
(see e.g.~\cite{Dumont:2014wha}), while the signal of {the charged} $SU(2)_L$  singlet is expected to be {yet} weaker. 
On the other hand, in the TII case one has an extra constraint  $|m_{\Delta^+}-m_{\Delta^{++}}|\lesssim 40\,\GeV$ from the electroweak precision data \cite{Chun:2012jw} which slightly tightens the bound.

Searches for doubly charged scalars are performed both by ATLAS and CMS ({see, e.g.,}~\cite{ATLAS:2012hi,Chatrchyan:2012ya}). 
These {analyses}  strongly depend on the assumed branching ratios and only leptonic decays are considered. 
{If, for instance, BR($h^{\pm\pm}\to l_1^\pm l_2^\pm$)=1 is assumed} for different pairs of $l_1 l_2$, the bounds vary from 200 to 450 GeV, {cf. \cite{Chatrchyan:2012ya} (note, however, that only the case of the type-II triplet was considered there).}
{To this end,} ref. \cite{ATLAS:2012hi} sets a bound of  410 GeV for the triplet case and up to 320 GeV for a doubly charged {$SU(2)_{L}$} singlet,
 again depending on the assumptions for the branching ratios.

In the case of the extended TII model, the branching ratios for the decay of {$\Delta^{++}$} into leptons are directly connected to the entries of the neutrino mass matrix. However, depending on the {shape} of $Y_{\Delta}$ (see \eq{Yukawa}), the leptonic $\Delta^{++}$ decays can be suppressed in favour of the decay into a pair of gauge bosons. Moreover, for nonzero mass splitting among the triplet components (which arises due to {the electroweak symmetry} breaking),  the cascade decays $\Delta^{\pm\pm}\to\Delta^{\pm} W^{\pm\ast}\to\Delta^{0} W^{\pm\ast} W^{\pm\ast}$ occur as well. For a complete ``decay phase diagram'' in the original TII model see {Ref.}~\cite{Melfo:2011nx}. 

Recently, the TII model extended by an extra Higgs doublet has been studied in \cite{Chen:2014qda} {with the conclusion} that the limit on the mass splittings is relaxed by the mixing and the dominant decay channel is {then} $\Delta^{\pm\pm}\to h^{\pm\ast}_{1,2} W^{\pm\ast}$. {The argument is} that a 5-$\sigma$ discovery of the doubly charged scalar is possible 
at LHC13 with an integrated luminosity of 40 fb$^{-1}$ if its mass is lower than 330 GeV. 
On the other hand, this conclusion does not apply to our TII model where
the form of the potential is constrained by the PQ symmetry (see \eq{TII-scalarpot}) and the size of the 
doublet-triplet mixing is always driven by the small triplet VEV (see \app{scalarspectrumII}).
 
As far as the ZB model is concerned, the authors of \cite{Herrero-Garcia:2014hfa} studied numerically the branching ratios of the doubly charged singlet $k^{++}$ and they {find a lower bound on its mass of about} 310 GeV in the case of {the} inverse neutrino hierarchy and 200 GeV for {the} normal hierarchy.

\subsection{Electroweak baryogenesis}
\label{EWbaryogenesis}

An essential ingredient for {the} electroweak baryogenesis (see, e.g.,~\cite{Morrissey:2012db} for a review) 
is a strong enough first-order phase transition. The finite-temperature effective potential must contain 
a {cubic term of the type {$|\phi|^3 T$} which is enhanced with respect to that available in the SM. 
This may be achieved if new light scalars with sizeable couplings to the Higgs doublet are present.}

There are two qualitatively different ways the new scalars can 
affect the electroweak phase transition: \emph{i)} by contributing directly to the effective potential through the 
evolution of their field values in the early universe, \emph{ii)} by enhancing the cubic term in the effective potential at the 
one-loop level without developing a VEV. 
{To this end, it is known, for instance} (see, e.g.,~\cite{Profumo:2007wc}), that 
the mechanism \emph{i)} works well when the new scalar developing a VEV is a gauge singlet $S$. 
In particular, for the mechanism to work, a crucial coupling to be present is the trilinear term $|H|^2 S$. 
Inspecting the different  potentials {studied in previous sections}, the only term that satisfies the above conditions (in the PQ-broken phase) is 
$\lambda_6 V_\sigma H_u^\dag {\bf \Delta} H_d$ 
in \eq{TII-scalarpot} of the extended TII model. On the other hand, this is likely not enough since {the}
neutrino masses (see \eq{neutrinomasses}) require this coupling to be very small. 
Hence, we are left with option \emph{ii)} for which the role of the inert scalar
running in the {loop} can be played by any of the two charged scalars $h^+$ and $k^{++}$ present in the radiative neutrino mass 
models, as well as the TII triplet, whose VEV is negligeable compared to the electroweak scale~\cite{AbdusSalam:2013eya}. 

{By denoting} the new charged state $X$ and by writing its interaction with the Higgs doublet as 
\begin{equation}
V = M_X^2 |X^2| + \lambda_{XH} |X|^2 |H|^2 + \ldots \, , 
\end{equation}
the contribution to the finite temperature effective potential 
due to the so-called daisy re-summation takes the form \cite{Morrissey:2012db}
\begin{equation}
\Delta V_{\rm{eff}} (\phi,T) \supset -\frac{n_X T}{12 \pi} \left[ \Pi_X (T) + M_X^2 + \frac{\lambda_{XH}}{2} \phi^2 \right]^{3/2} \, , 
\end{equation}
where $n_X$ is the number of degrees of freedom and 
$\Pi_X (T) = \kappa T^2$ is the thermal mass of $X$, 
with $\kappa$ being a function {of the} scalar and gauge couplings. 
Hence, in order to maximize the contribution to the cubic term 
one needs a {significant} portal coupling $\lambda_{XH}$ and/or a cancellation between the thermal mass and the 
potential mass parameter. While it is known that a strong first-order phase transition 
can be achieved in the case of a second inert scalar doublet~\cite{Chowdhury:2011ga} or an additional triplet~\cite{AbdusSalam:2013eya}, a detailed study of its feasibility in the three setups here discussed is left to further investigation. 

Finally, let us comment on the additional sources of CP violation, {another necessary} ingredient for a successful baryogenesis. 
By taking advantage of the results of Ref.~\cite{Haber:2012np}, it is readily shown that none of the potentials considered in our setups 
lead to spontaneous CP violation at {the} tree level. 
This statement is verified by a direct minimization of the relevant scalar potentials in the presence of complex VEVs. 
Hence, an extension of the {minimal setups here discussed } is required if the new source of CP violation has to come from the scalar potential. The simplest option is adding a SM singlet \cite{Geng:1988ty}
or to consider a PQ-extended three-Higgs-doublet model \cite{He:1988dm}. 
Another intriguing possibility is {to identify the source of the extra CP violation with the very $\overline{\theta}_{\rm{QCD}}$-term} 
which, in the early universe, has not yet relaxed to its minimum. Such a scenario can be realized in the context of the so-called 
cold electroweak baryogenesis where the electroweak phase transition is delayed to temperatures 
{$T\lesssim \Lambda_{\rm{QCD}}$ \cite{Servant:2014bla}. Further scrutiny on the matter is called for.

\section{Conclusions}\label{conclusions}

Inspired by the present-day evidence for physics beyond the standard electroweak model we have {considered} three simple setups that minimally extend the scalar sector of the SM. They feature massive {Majorana} neutrinos together with an invisible axion, thus interconnecting neutrino masses, dark matter and the strong CP problem. {In all cases,} the presence of the PQ symmetry strongly constrains the {relevant scalar potentials}, replacing the role of the ad hoc discrete symmetries originally invoked in some of the models.

The Higgs sectors of the extended setups feature in all cases two Higgs doublets and one complex scalar singlet.  Two widely different physical scales are connected in {these} setups: the PQ and {the} electroweak scale. It is noteworthy that the neutrino mass scale is {fully compatible} with the requirement of naturalness and stability of the scalar masses. The needed decoupling (ultraweak) limit of the PQ singlet scalar is shown to be technically natural and allows for a plethora of physical scalar states at the TeV scale, {within} the reach of the collider searches. We commented upon the possible destabilizing role of gravity, but we {feel the arguments do not compel us} to abandon such a scenario.
 
The {same extensions} of the SM scalar sector allow for an improvement on the electroweak vacuum stability {issue} and, at the same time, they may trigger
a {strong-enough} first-order electroweak phase transition {to support the } electroweak baryogenesis as the origin of the observed baryon asymmetry.
{Alas}, the minimal scalar potentials here considered do not allow for additional sources of CP violation. The matter will be the subject of a further investigation.

Finally, it is intriguing that the hierarchy between the PQ and {the} electroweak scales may naturally arise as a consequence of the breaking 
of {the} classical scale invariance \'a la Coleman-Weinberg \cite{Allison:2014zya,Allison:2014hna}. Due to the absence of scalar trilinear interactions which characterizes  the neutrino-invisible axion models here discussed, they are naturally and readily embedded into such a context (the extended Poincar\'e invariance of the action in the ultraweak limit is there replaced by a custodial shift symmetry of the singlet field). Again, this scenario will be scrutinized in a future work.

\section*{Acknowledgments}

\noindent

S.B.~acknowledges partial support by the italian MIUR grant
no.~2010YJ2NYW001 and by the EU Marie Curie ITN UNILHC grant no.~PITN-GA-2009-237920. 
The work of L.D.L.~is supported by the Marie Curie CIG program, project number PCIG13-GA-2013-618439. 
L.D.L.~is grateful to SISSA for hospitality and partial support during the development of this project. 
The work of H.K.~is supported by the Grant Agency of the Czech Technical University in Prague, grant No.~SGS13/217/OHK4/3T/14. {The work of M.M. is supported by the Marie-Curie Career Integration Grant within the 7th European Community Framework Programme
FP7-PEOPLE-2011-CIG, contract number PCIG10-GA-2011-303565 and by the by the Foundation for support of science and research ``Neuron''.}

\appendix

\section{{Scalar spectrum of the TII model}}\label{scalarspectrumII}

In this appendix we detail the scalar spectrum of the extended Type-II {seesaw} model. 
First, we consider the case where all the electroweak breaking VEVs are neglected, 
thus providing a very simple description of the spectrum in the PQ-broken phase. 
Later we discuss the general case by taking advantage of the large VEVs hierarchies.

\subsection{$v_{u,d} = v_\Delta = 0$ case}
\label{vudvDeltaeq0}

With all the electroweak VEVs switched off ($v_{u,d} = v_\Delta = 0$), 
we only retain the PQ-symmetry-breaking VEV $V_\sigma$ {in \eq{expsigma}},  yielding the 
{stationarity} equation
\begin{equation}
\label{statcondvsigmanoew}
0 = \frac{\partial \vev{V_{\rm{TII}}} }{\partial V_\sigma} = 
2 V_{\sigma } 
\left(
2 \lambda _3 V_{\sigma }^2
- \mu _3^2
\right) \, .
\end{equation}
Expanding the scalar potential up to {the} second order in the dynamical fields
and using (\ref{statcondvsigmanoew}), we obtain the following spectrum 
classified according to the unbroken SM symmetry: 

\noindent {{{\em i)} A real} scalar SM singlet $\sigma^0$:}
\begin{equation}
\label{M2singlets}
M^2_{\sigma^0} = 4 \lambda _3 V_{\sigma }^2 \, .
\end{equation}
{{{\em ii)} A real} pseudoscalar SM singlet $\eta_\sigma^0$:}
\begin{equation}
\label{M2singletps}
M^2_{\eta_\sigma^0} = 0 \, ,
\end{equation}
which is the {zero-mass mode of the PQ-breaking field} corresponding to the axion.\\ 
{{{\em iii)} A complex} triplet $\Delta$:}
\begin{equation}
\label{M2triplet}
M^2_{\Delta} = \lambda _{\text{$\Delta $3}} V_{\sigma }^2 +\mu _{\Delta}^2 \, .
\end{equation}
{{\em iv)} Complex doublets $H_u$ and $H_d$:}
\begin{equation}
\label{M2doublets}
M^2_{H} = 
\left(
\begin{array}{cc}
 \lambda _{13} V_{\sigma }^2 -\mu _1^2 & \lambda _5 V_{\sigma }^2 \\
 \lambda _5 V_{\sigma }^2 & \lambda _{23} V_{\sigma }^2-\mu _2^2 \\
\end{array}
\right) 
\, ;
\end{equation}
{here $M^2_{H}$ is written in} the $(H_u^*, H_d)$ basis (column indices).
\eq{M2doublets} is diagonalized by an orthogonal transformation 
\begin{equation}
\label{orthogtransf}
\left(
\begin{array}{c}
\hat{H}_u^* \\
\hat{H}_d 
\end{array}
\right) 
=
\left(
\begin{array}{cc}
\cos{\alpha} & \sin{\alpha} \\
-\sin{\alpha} & \cos{\alpha} 
\end{array}
\right) 
\left(
\begin{array}{c}
H_u^* \\
H_d 
\end{array}
\right) \, , 
\end{equation}
where 
\begin{equation}
\label{tan2alpha}
\tan 2\alpha = \frac{2 \lambda _5 V_{\sigma }^2}{\left(\lambda _{13} - \lambda _{23} \right) V_{\sigma }^2 -\mu _1^2  + \mu _2^2} \, .
\end{equation}
The corresponding eigenvalues then read
\begin{multline}
\label{eigenval}
2 M^2_{u,d} =  \left( \lambda _{13} + \lambda _{23} \right) V_{\sigma }^2  -\mu _1^2 -\mu _2^2  
\\ \pm \sqrt{ \left( \left(\lambda _{13} - \lambda _{23} \right) V_{\sigma }^2 -\mu _1^2  + \mu _2^2 \right)^2 + 4 \lambda _5^2 V_{\sigma }^4 }  \, .
\end{multline}

\subsection{$v_{u,d,\Delta} \neq 0$ case}
\label{vudvDeltaneq0}

By plugging \eqs{expHu}{expsigma} into the expression of the scalar potential in 
\eq{TII-scalarpot}, we obtain the {stationarity equations in the form}
\begin{widetext}
\begin{align}
\label{statcondv1}
0 & = \frac{\partial \vev{V_{\rm{TII}}} }{\partial v_u} =
2v_u \left(2 \lambda _1 v_u^2 + \lambda _{12} v_d^2 + \lambda _{13} V_{\sigma }^2 
+  \lambda _{\text{$\Delta $1}} v_{\Delta }^2
- \mu _1^2 \right)+
   2 \lambda _6 v_{\Delta } V_{\sigma } v_d - 2 \lambda _5 v_d V_{\sigma }^2 \, , \\
\label{statcondv2}
0 & = \frac{\partial \vev{V_{\rm{TII}}} }{\partial v_d} =
2v_d \left(
2 \lambda _2 v_d^2 
+\lambda _{12} v_u^2
+ \lambda _{23} V_{\sigma }^2
+ (\lambda _{\text{$\Delta $2}} + \lambda _8) v_{\Delta }^2 
- \mu _2^2
\right)
+2 \lambda _6 v_{\Delta } V_{\sigma } v_u
-2 \lambda _5 v_u V_{\sigma}^2 \, ,
\\
\label{statcondvsigma}
0 & = \frac{\partial \vev{V_{\rm{TII}}} }{\partial V_\sigma} = 
2 V_{\sigma } 
\left(
2 \lambda _3 V_{\sigma }^2
+ \lambda _{13} v_u^2
+ \lambda _{23} v_d^2
-2 \lambda _5 v_u v_d
+ \lambda _{\text{$\Delta $3}} v_{\Delta }^2
- \mu _3^2
\right)+2\lambda _6 v_{\Delta }
   v_u v_d \, ,
\\
\label{statcondvDelta}
0 & = \frac{\partial \vev{V_{\rm{TII}}} }{\partial v_\Delta} =
2 v_{\Delta } 
\left(
2 \left( \lambda _{\text{$\Delta $4}} + \lambda _9\right) v_{\Delta }^2
+ \lambda_{\text{$\Delta $1}} v_u^2
+ (\lambda _{\text{$\Delta $2}}
+ \lambda _8) v_d^2
+  \lambda _{\text{$\Delta $3}} V_{\sigma }^2
+ \mu _{\Delta}^2
\right)
+2\lambda _6 V_{\sigma } v_u v_d \, .
\end{align}   

The spectrum is then obtained by expanding the scalar potential up to {the} second order in the fields of \eqs{expHu}{expsigma},
{around the vacuum configuration given by the above stationarity} equations. This yields: 
\\[1ex]
{\em i)} Neutral scalars $(h^0_u, h^0_d, \sigma^0, \delta^0)$:
\begin{multline}
\label{m2s}
M^2_{\text{S}} = 
\left(
\begin{array}{cc}
 4 \lambda _1 v_u^2+ \left(\lambda _5 V_{\sigma } - \lambda _6 v_{\Delta } \right) V_{\sigma } v_d / v_u
   &  \left(\lambda _6 v_{\Delta } - \lambda _5 V_{\sigma } \right) V_{\sigma } +2 \lambda _{12} v_u v_d  \\
  \left(\lambda _6 v_{\Delta } - \lambda _5 V_{\sigma } \right) V_{\sigma } 
  +2 \lambda _{12} v_u v_d  
  & 4 \lambda _2 v_d^2+ \left(\lambda _5 V_{\sigma } -\lambda _6 v_{\Delta } \right)V_{\sigma } v_u / v_d \\
 -2 \lambda _5 V_{\sigma } v_d +\lambda _6 v_{\Delta } v_d +2 \lambda _{13} V_{\sigma } v_u  
 & -2 \lambda _5 V_{\sigma } v_u +\lambda _6 v_{\Delta } v_u +2 \lambda _{23} V_{\sigma } v_d  \\
 \lambda _6 V_{\sigma } v_d +2 \lambda _{\text{$\Delta $1}} v_{\Delta } v_u  & \lambda _6 V_{\sigma } v_u 
 +2 \left(\lambda _8+\lambda _{\text{$\Delta $2}}\right) v_{\Delta } v_d  \\
\end{array} 
\right. \\
\left.
\begin{array}{cccc}
 -2 \lambda _5 V_{\sigma } v_d + \lambda _6 v_{\Delta } v_d +2 \lambda _{13} V_{\sigma } v_u  
 & \lambda _6 V_{\sigma } v_d + 2 \lambda _{\text{$\Delta $1}} v_{\Delta } v_u  \\
 -2 \lambda _5 V_{\sigma } v_u + \lambda _6 v_{\Delta } v_u +2 \lambda _{23} V_{\sigma } v_d 
 & \lambda _6 V_{\sigma } v_u +2 \left(\lambda _8+\lambda _{\text{$\Delta $2}}\right) v_{\Delta } v_d  \\
  4 \lambda_3 V_{\sigma }^2 - \lambda _6 v_{\Delta } v_u v_d / V_{\sigma } 
  & \lambda _6 v_u v_d +2 \lambda _{\text{$\Delta $3}} v_{\Delta } V_{\sigma } \\
  \lambda _6 v_u v_d +2 \lambda _{\text{$\Delta $3}} v_{\Delta } V_{\sigma } 
  & 4 \left(\lambda _9+\lambda_{\text{$\Delta $4}}\right) v_{\Delta }^2 - \lambda _6 V_{\sigma } v_d v_d / v_{\Delta } \\
\end{array}
\right) \, ,
\end{multline}
where $\text{Rank} \ M^2_{\text{S}} = 4$. The exact form of the eigenvalues is quite cumbersome. However, the {required} hierarchy $v_\Delta\ll v_u,v_d\ll V_\sigma$ allows us to compute the eigenvalues perturbatively. Taking into account the scaling of the couplings in \eq{lambdahierarchies}, we define $\lambda_6 \equiv c_6 \frac{v_\Delta}{V_\sigma}$, $\lambda_{5} \equiv c_{5} \frac{v^2}{V_\sigma^2}$, $\lambda_{i3} \equiv c_{i3} \frac{v^2}{V_\sigma^2}$ with $c_5, c_6, c_{i3}$ being $\mathcal{O}(1)$ coefficients. Hence, the leading contribution to the neutral scalar mass matrix  ({given by the} terms $\gtrsim \mathcal{O}(v^2)$) reads
$$
M^{2(\text{LO})}_{\text{S}} = 
\left(
\begin{array}{cccc}
 4 \lambda _1 v_u^2+ c_5\frac{v_d v^2}{v_u} &  c_5 v^2 +2 \lambda_{12} v_u v_d  & 0 & 0 \\
  c_5 v^2 +2 \lambda _{12} v_u v_d  & 4 \lambda _2 v_d^2+ c_5 \frac{v_u v^2}{v_d} &  0 & 0 \\
 0 & 0 &  4 \lambda_3 V_{\sigma }^2 & 0 \\
0  & 0  & 0  & - c_6 v_u v_d\\
\end{array}
\right)\, .\\
$$
It is now {clear} that, at the leading order in the VEV ratio expansion, the eigenvalues of the scalar mass matrix read
\begin{equation}\label{EVS}
\{\mathcal{O}(v^2), \, \mathcal{O}(v^2),\, 4\lambda_3 V_\sigma^2, \, -v_u v_d c_6\}
\end{equation}
and that there is no mixing of the singlet and triplet fields with the $SU(2)_L$ doublets. The first NLO corrections to the mass matrix are {of the order of} $v_\Delta v$, which implies, {for instance,}~that the mixing between the doublet and {the} triplet components is of {the order of} $v_\Delta/v$.  One further finds that the first corrections to the eigenvalues are only of {the order of} $v_\Delta^2$ and that the mixing with the singlet component is {of the order of} $v/V_\sigma$. For large $\tan\beta=v_u/v_d$ the two doublet eigenvalues are approximately $4\lambda_1 v^2$ and $c_5 v^2 \tan\beta$, while for the mixing angle $\alpha$ one obtains $\tan\alpha \approx \cot\beta \ll 1$. In this limit the lightest doublet scalar behaves as the standard model Higgs boson.  
\\[1ex]
{{\em ii)}} Neutral pseudo-scalars $(\eta^0_u, \eta^0_d, \eta_\sigma^0, \eta_\delta^0)$: 
\begin{equation}
\label{m2ps}
M^2_{\text{PS}} = 
\left(
\begin{array}{cccc}
 \left(\lambda _5 V_{\sigma } - \lambda _6 v_{\Delta } \right) V_{\sigma } v_d / v_u
 &  \left(\lambda _5 V_{\sigma } + \lambda _6 v_{\Delta } \right) V_{\sigma }
   & \left(2 \lambda_5 V_{\sigma } +\lambda _6 v_{\Delta } \right) v_d
   & - \lambda _6 V_{\sigma } v_d \\
 \left(\lambda _5 V_{\sigma } + \lambda _6 v_{\Delta } \right) V_{\sigma }
 & \left(\lambda _5 V_{\sigma } - \lambda _6 v_{\Delta } \right) V_{\sigma } v_u / v_d
 & \left(2 \lambda_5 V_{\sigma } - \lambda _6 v_{\Delta } \right) v_u
 & \lambda _6 V_{\sigma } v_u  \\
 \left(2 \lambda _5 V_{\sigma } +\lambda _6 v_{\Delta } \right) v_d 
 & \left(2 \lambda_5 V_{\sigma } - \lambda _6 v_{\Delta } \right) v_u 
 & 4 \lambda _5  v_u v_d - \lambda_6 v_{\Delta } v_u v_d / V_{\sigma } 
 & \lambda _6 v_u v_d \\
 - \lambda _6 V_{\sigma } v_d 
 & \lambda _6 V_{\sigma } v_u 
 & \lambda _6 v_u v_d 
 & - \lambda _6 V_{\sigma } v_u v_d / v_{\Delta } \\
\end{array}
\right) \, 
\end{equation}
{is a $\text{Rank}=2$ matrix which implies} the existence of two zero modes, {one of them being} the would-be Goldstone mode associated with the $Z$ boson and the other corresponding to the axion that acquires mass by non-perturbative QCD effects. Even though the eigenvalues can be given in a closed {form}, it is sufficient to report the LO result
\begin{equation}\label{EVPS}
\left\{0, \frac{v^4}{v_u v_d}c_5, 0, -v_u v_d c_6\right\},
\end{equation}
where the {entries correspond, consecutively,} to the pair of $SU(2)_{L}$ (mostly) doublet components, the singlet and the triplet. The zeros are exact {(at the perturbative level), while the other entries receive corrections at most of the order of $v_\Delta^2$}. The mixing among the doublet and triplet components is again found to be of {the order of} $v_\Delta/v$.
\\[1ex]
{{\em iii)}} Singly-charged scalars: $(h^+_u, h^+_d, \delta^+)$ 
 \begin{equation}
\label{m2scsc}
M^2_{+} = 
\left(
\begin{array}{ccc}
 \lambda_4 v_d^2 
 + \lambda _7 v_{\Delta }^2
 + \left( \lambda _5 V_{\sigma } 
 -\lambda _6 v_{\Delta }  \right) V_{\sigma } v_d / v_u 
 & \lambda _5 V_{\sigma }^2 
 +\lambda _4 v_u v_d &
 \frac{\lambda _7 v_{\Delta } v_u -\lambda _6V_{\sigma } v_d }{\sqrt{2}} \\
 \lambda _5 V_{\sigma }^2
 +\lambda _4 v_u v_d  
 & \lambda_4 v_u^2 -\lambda _8 v_{\Delta }^2 + \left(\lambda _5 V_{\sigma } -\lambda _6 v_{\Delta } \right)V_{\sigma } 
 v_u / v_d
 &  \frac{\lambda _6 V_{\sigma } v_u 
 +\lambda _8 v_{\Delta } v_d }{\sqrt{2}} \\
 \frac{\lambda _7 v_{\Delta } v_u - \lambda _6 V_{\sigma } v_d }{\sqrt{2}} 
 & \frac{\lambda_6 V_{\sigma } v_u +\lambda _8 v_{\Delta } v_d }{\sqrt{2}} 
 & 
 \frac{ \lambda _7 v_u^2 - \lambda _8 v_d^2 }{2} - \lambda _6 V_{\sigma } v_d  v_u / v_{\Delta }  \\
\end{array}
\right) \, 
\end{equation}
{is again of Rank 2, which is} related to the existence of {a would-be Goldstone mode} associated to the $W$ boson. 
In analogy with the PS case, the eigenvalues read at LO
\begin{equation}\label{EVChS}
\left\{0,\lambda_4 v^2 + c_5 \frac{v^4}{v_u v_d}, -c_6 v_u v_d + \frac{1}{2}\left(\lambda_7 v_u^2 + \lambda_8 v_d^2\right)\right\},\end{equation}
and the mixing of the doublet and triplet components is suppressed by {the} $v_\Delta/v$ ratio.
\\[1ex]
{{\em iv)}} Doubly-charged scalar $\delta^{++}$: 
\begin{equation}
\label{m2dcsc}
M^2_{++} = 
\lambda _7 v_u^2 
- \lambda _8 v_d^2  
- 2 \lambda _9 v_{\Delta }^2
-  \lambda _6 v_u v_d V_{\sigma }/ v_{\Delta } \approx \lambda _7 v_u^2 
- \lambda _8 v_d^2  
-  c_6 v_u v_d \, .
\end{equation}
{By comparing} \eqref{m2dcsc} with \eqref{EVS}, \eqref{EVPS} and \eqref{EVChS} one recognizes the weak mass splitting among the triplet components induced, at the leading order, by the $\lambda_7$ and $\lambda_8$ terms.
\end{widetext}

%
\bibliographystyle{utphysmod}
\bibliography{nupq}

\providecommand{\href}[2]{#2}\begingroup\raggedright\begin{thebibliography}{10}

\bibitem{Farina:2013mla}
M.~Farina, D.~Pappadopulo, and A.~Strumia, ``{A modified naturalness principle
  and its experimental tests}'',
  \href{http://dx.doi.org/10.1007/JHEP08(2013)022}{\blue JHEP {\bfseries 1308}
  (2013) 022},
\href{http://arxiv.org/abs/1303.7244}{{\blue arXiv:1303.7244 [hep-ph]}}.

\bibitem{EliasMiro:2012ay}
J.~Elias-Miro, J.~R. Espinosa, G.~F. Giudice, H.~M. Lee, and A.~Strumia,
  ``{Stabilization of the Electroweak Vacuum by a Scalar Threshold Effect}'',
  \href{http://dx.doi.org/10.1007/JHEP06(2012)031}{\blue JHEP {\bfseries 1206}
  (2012) 031},
\href{http://arxiv.org/abs/1203.0237}{{\blue arXiv:1203.0237 [hep-ph]}}.

\bibitem{Schechter:1980gr}
J.~Schechter and J.~Valle, ``{Neutrino Masses in SU(2) x U(1) Theories}'',
\href{http://dx.doi.org/10.1103/PhysRevD.22.2227}{\blue Phys.Rev. {\bfseries
  D22} (1980) 2227}.

\bibitem{Cheng:1980qt}
T.~Cheng and L.-F. Li, ``{Neutrino Masses, Mixings and Oscillations in SU(2) x
  U(1) Models of Electroweak Interactions}'',
\href{http://dx.doi.org/10.1103/PhysRevD.22.2860}{\blue Phys.Rev. {\bfseries
  D22} (1980) 2860}.

\bibitem{Lazarides:1980nt}
G.~Lazarides, Q.~Shafi, and C.~Wetterich, ``{Proton Lifetime and Fermion Masses
  in an SO(10) Model}'',
\href{http://dx.doi.org/10.1016/0550-3213(81)90354-0}{\blue Nucl.Phys.
  {\bfseries B181} (1981) 287--300}.

\bibitem{Mohapatra:1980yp}
R.~N. Mohapatra and G.~Senjanovic, ``{Neutrino Masses and Mixings in Gauge
  Models with Spontaneous Parity Violation}'',
\href{http://dx.doi.org/10.1103/PhysRevD.23.165}{\blue Phys.Rev. {\bfseries
  D23} (1981) 165}.

\bibitem{Wetterich:1981bx}
C.~Wetterich, ``{Neutrino Masses and the Scale of B-L Violation}'',
\href{http://dx.doi.org/10.1016/0550-3213(81)90279-0}{\blue Nucl.Phys.
  {\bfseries B187} (1981) 343}.

\bibitem{Zee:1980ai}
A.~Zee, ``{A Theory of Lepton Number Violation, Neutrino Majorana Mass, and
  Oscillation}'',
\href{http://dx.doi.org/10.1016/0370-2693(80)90349-4}{\blue Phys.Lett.
  {\bfseries B93} (1980) 389}.

\bibitem{Ma:1998dn}
E.~Ma, ``{Pathways to naturally small neutrino masses}'',
  \href{http://dx.doi.org/10.1103/PhysRevLett.81.1171}{\blue Phys.Rev.Lett.
  {\bfseries 81} (1998) 1171--1174},
\href{http://arxiv.org/abs/hep-ph/9805219}{{\blue arXiv:hep-ph/9805219
  [hep-ph]}}.

\bibitem{Babu:2001ex}
K.~Babu and C.~N. Leung, ``{Classification of effective neutrino mass
  operators}'', \href{http://dx.doi.org/10.1016/S0550-3213(01)00504-1}{\blue
  Nucl.Phys. {\bfseries B619} (2001) 667--689},
\href{http://arxiv.org/abs/hep-ph/0106054}{{\blue arXiv:hep-ph/0106054
  [hep-ph]}}.

\bibitem{Bonnet:2012kz}
F.~Bonnet, M.~Hirsch, T.~Ota, and W.~Winter, ``{Systematic study of the d=5
  Weinberg operator at one-loop order}'',
  \href{http://dx.doi.org/10.1007/JHEP07(2012)153}{\blue JHEP {\bfseries 1207}
  (2012) 153},
\href{http://arxiv.org/abs/1204.5862}{{\blue arXiv:1204.5862 [hep-ph]}}.

\bibitem{Wolfenstein:1980sy}
L.~Wolfenstein, ``{A Theoretical Pattern for Neutrino Oscillations}'',
\href{http://dx.doi.org/10.1016/0550-3213(80)90004-8}{\blue Nucl.Phys.
  {\bfseries B175} (1980) 93}.

\bibitem{Koide:2001xy}
Y.~Koide, ``{Can the Zee model explain the observed neutrino data?}'',
  \href{http://dx.doi.org/10.1103/PhysRevD.64.077301}{\blue Phys.Rev.
  {\bfseries D64} (2001) 077301},
\href{http://arxiv.org/abs/hep-ph/0104226}{{\blue arXiv:hep-ph/0104226
  [hep-ph]}}.

\bibitem{Frampton:2001eu}
P.~H. Frampton, M.~C. Oh, and T.~Yoshikawa, ``{Zee model confronts SNO data}'',
  \href{http://dx.doi.org/10.1103/PhysRevD.65.073014}{\blue Phys.Rev.
  {\bfseries D65} (2002) 073014},
\href{http://arxiv.org/abs/hep-ph/0110300}{{\blue arXiv:hep-ph/0110300
  [hep-ph]}}.

\bibitem{He:2003ih}
X.-G. He, ``{Is the Zee model neutrino mass matrix ruled out?}'',
  \href{http://dx.doi.org/10.1140/epjc/s2004-01669-8}{\blue Eur.Phys.J.
  {\bfseries C34} (2004) 371--376},
\href{http://arxiv.org/abs/hep-ph/0307172}{{\blue arXiv:hep-ph/0307172
  [hep-ph]}}.

\bibitem{Babu:2013pma}
K.~Babu and J.~Julio, ``{Predictive Model of Radiative Neutrino Masses}'',
  \href{http://dx.doi.org/10.1103/PhysRevD.89.053004}{\blue Phys.Rev.
  {\bfseries D89} no.~5, (2014) 053004},
\href{http://arxiv.org/abs/1310.0303}{{\blue arXiv:1310.0303 [hep-ph]}}.

\bibitem{Zee:1985id}
A.~Zee, ``{Quantum Numbers of Majorana Neutrino Masses}'',
\href{http://dx.doi.org/10.1016/0550-3213(86)90475-X}{\blue Nucl.Phys.
  {\bfseries B264} (1986) 99}.

\bibitem{Babu:1988ki}
K.~Babu, ``{Model of 'Calculable' Majorana Neutrino Masses}'',
\href{http://dx.doi.org/10.1016/0370-2693(88)91584-5}{\blue Phys.Lett.
  {\bfseries B203} (1988) 132}.

\bibitem{Sierra:2014rxa}
D.~A. Sierra, A.~Degee, L.~Dorame, and M.~Hirsch, ``{Systematic classification
  of two-loop realizations of the Weinberg operator}'',
\href{http://arxiv.org/abs/1411.7038}{{\blue arXiv:1411.7038 [hep-ph]}}.

\bibitem{Herrero-Garcia:2014hfa}
J.~Herrero-Garcia, M.~Nebot, N.~Rius, and A.~Santamaria, ``{The Zee-Babu Model
  revisited in the light of new data}'',
  \href{http://dx.doi.org/10.1016/j.nuclphysb.2014.06.001}{\blue Nucl.Phys.
  {\bfseries B885} (2014) 542--570},
\href{http://arxiv.org/abs/1402.4491}{{\blue arXiv:1402.4491 [hep-ph]}}.

\bibitem{Schmidt:2014zoa}
D.~Schmidt, T.~Schwetz, and H.~Zhang, ``{Status of the ZeeÐBabu model for
  neutrino mass and possible tests at a like-sign linear collider}'',
  \href{http://dx.doi.org/10.1016/j.nuclphysb.2014.05.024}{\blue Nucl.Phys.
  {\bfseries B885} (2014) 524--541},
\href{http://arxiv.org/abs/1402.2251}{{\blue arXiv:1402.2251 [hep-ph]}}.

\bibitem{Peccei:1977hh}
R.~Peccei and H.~R. Quinn, ``{CP Conservation in the Presence of Instantons}'',
\href{http://dx.doi.org/10.1103/PhysRevLett.38.1440}{\blue Phys.Rev.Lett.
  {\bfseries 38} (1977) 1440--1443}.

\bibitem{Peccei:1977ur}
R.~Peccei and H.~R. Quinn, ``{Constraints Imposed by CP Conservation in the
  Presence of Instantons}'',
\href{http://dx.doi.org/10.1103/PhysRevD.16.1791}{\blue Phys.Rev. {\bfseries
  D16} (1977) 1791--1797}.

\bibitem{Weinberg:1977ma}
S.~Weinberg, ``{A New Light Boson?}'',
\href{http://dx.doi.org/10.1103/PhysRevLett.40.223}{\blue Phys.Rev.Lett.
  {\bfseries 40} (1978) 223--226}.

\bibitem{Wilczek:1977pj}
F.~Wilczek, ``{Problem of Strong p and t Invariance in the Presence of
  Instantons}'',
\href{http://dx.doi.org/10.1103/PhysRevLett.40.279}{\blue Phys.Rev.Lett.
  {\bfseries 40} (1978) 279--282}.

\bibitem{Weinberg:1975ui}
S.~Weinberg, ``{The U(1) Problem}'',
\href{http://dx.doi.org/10.1103/PhysRevD.11.3583}{\blue Phys.Rev. {\bfseries
  D11} (1975) 3583--3593}.

\bibitem{'tHooft:1976fv}
G.~'t~Hooft, ``{Computation of the Quantum Effects Due to a Four-Dimensional
  Pseudoparticle}'',
\href{http://dx.doi.org/10.1103/PhysRevD.18.2199.3,
  10.1103/PhysRevD.14.3432}{\blue Phys.Rev. {\bfseries D14} (1976) 3432--3450}.

\bibitem{Callan:1976je}
J.~Callan, Curtis~G., R.~Dashen, and D.~J. Gross, ``{The Structure of the Gauge
  Theory Vacuum}'',
\href{http://dx.doi.org/10.1016/0370-2693(76)90277-X}{\blue Phys.Lett.
  {\bfseries B63} (1976) 334--340}.

\bibitem{Jackiw:1976pf}
R.~Jackiw and C.~Rebbi, ``{Vacuum Periodicity in a Yang-Mills Quantum
  Theory}'',
\href{http://dx.doi.org/10.1103/PhysRevLett.37.172}{\blue Phys.Rev.Lett.
  {\bfseries 37} (1976) 172--175}.

\bibitem{'tHooft:1986nc}
G.~'t~Hooft, ``{How Instantons Solve the U(1) Problem}'',
\href{http://dx.doi.org/10.1016/0370-1573(86)90117-1}{\blue Phys.Rept.
  {\bfseries 142} (1986) 357--387}.

\bibitem{Kim:2008hd}
J.~E. Kim and G.~Carosi, ``{Axions and the Strong CP Problem}'',
  \href{http://dx.doi.org/10.1103/RevModPhys.82.557}{\blue Rev.Mod.Phys.
  {\bfseries 82} (2010) 557--602},
\href{http://arxiv.org/abs/0807.3125}{{\blue arXiv:0807.3125 [hep-ph]}}.

\bibitem{Mohapatra:1982tc}
R.~N. Mohapatra and G.~Senjanovic, ``{The Superlight Axion and Neutrino
  Masses}'',
\href{http://dx.doi.org/10.1007/BF01577819}{\blue Z.Phys. {\bfseries C17}
  (1983) 53--56}.

\bibitem{Shafi:1984ek}
Q.~Shafi and F.~Stecker, ``{Implications of a Class of Grand Unified Theories
  for Large Scale Structure in the Universe}'',
\href{http://dx.doi.org/10.1103/PhysRevLett.53.1292}{\blue Phys.Rev.Lett.
  {\bfseries 53} (1984) 1292}.

\bibitem{Langacker:1986rj}
P.~Langacker, R.~Peccei, and T.~Yanagida, ``{Invisible Axions and Light
  Neutrinos: Are They Connected?}'',
\href{http://dx.doi.org/10.1142/S0217732386000683}{\blue Mod.Phys.Lett.
  {\bfseries A1} (1986) 541}.

\bibitem{Shin:1987xc}
M.~Shin, ``{Light Neutrino Masses and Strong {CP} Problem}'',
\href{http://dx.doi.org/10.1103/PhysRevLett.59.2515}{\blue Phys.Rev.Lett.
  {\bfseries 59} (1987) 2515}.

\bibitem{He:1988dm}
X.~He and R.~Volkas, ``{Models Featuring Spontaneous {CP} Violation: An
  Invisible Axion and Light Neutrino Masses}'',
\href{http://dx.doi.org/10.1016/0370-2693(88)90427-3}{\blue Phys.Lett.
  {\bfseries B208} (1988) 261}.

\bibitem{Geng:1988nc}
C.~Geng and J.~N. Ng, ``{Flavor Connections and Neutrino Mass Hierarchy
  Invariant Invisible Axion Models Without Domain Wall Problem}'',
\href{http://dx.doi.org/10.1103/PhysRevD.39.1449}{\blue Phys.Rev. {\bfseries
  D39} (1989) 1449}.

\bibitem{Berezhiani:1989fp}
Z.~Berezhiani and M.~Y. Khlopov, ``{Cosmology of Spontaneously Broken Gauge
  Family Symmetry with axion solution of strong CP-problem}'',
\href{http://dx.doi.org/10.1007/BF01570798}{\blue Z.Phys. {\bfseries C49}
  (1991) 73--78}.

\bibitem{Bertolini:1990vz}
S.~Bertolini and A.~Santamaria, ``{The Strong CP problem and the solar neutrino
  puzzle: Are they related?}'',
\href{http://dx.doi.org/10.1016/0550-3213(91)90467-C}{\blue Nucl.Phys.
  {\bfseries B357} (1991) 222--240}.

\bibitem{Arason:1990sg}
H.~Arason, P.~Ramond, and B.~Wright, ``{A Standard model extension with
  neutrino masses and an invisible axion}'',
\href{http://dx.doi.org/10.1103/PhysRevD.43.2337}{\blue Phys.Rev. {\bfseries
  D43} (1991) 2337--2350}.

\bibitem{Ma:2001ac}
E.~Ma, ``{Making neutrinos massive with an axion in supersymmetry}'',
  \href{http://dx.doi.org/10.1016/S0370-2693(01)00787-0}{\blue Phys.Lett.
  {\bfseries B514} (2001) 330--334},
\href{http://arxiv.org/abs/hep-ph/0102008}{{\blue arXiv:hep-ph/0102008
  [hep-ph]}}.

\bibitem{Dias:2005dn}
A.~G. Dias and V.~Pleitez, ``{The Invisible axion and neutrino masses}'',
  \href{http://dx.doi.org/10.1103/PhysRevD.73.017701}{\blue Phys.Rev.
  {\bfseries D73} (2006) 017701},
\href{http://arxiv.org/abs/hep-ph/0511104}{{\blue arXiv:hep-ph/0511104
  [hep-ph]}}.

\bibitem{Ma:2011rp}
E.~Ma, ``{Supersymmetric Axion-Neutrino Model with a Higgs Hybrid}'',
  \href{http://dx.doi.org/10.1142/S0217751X12500595}{\blue Int.J.Mod.Phys.
  {\bfseries A27} (2012) 1250059},
\href{http://arxiv.org/abs/1112.1367}{{\blue arXiv:1112.1367 [hep-ph]}}.

\bibitem{Chen:2012baa}
C.-S. Chen and L.-H. Tsai, ``{Peccei-Quinn symmetry as the origin of Dirac
  Neutrino Masses}'', \href{http://dx.doi.org/10.1103/PhysRevD.88.055015}{\blue
  Phys.Rev. {\bfseries D88} no.~5, (2013) 055015},
\href{http://arxiv.org/abs/1210.6264}{{\blue arXiv:1210.6264 [hep-ph]}}.

\bibitem{Park:2014qha}
W.-I. Park, ``{PQ-symmetry for a small Dirac neutrino mass, dark radiation and
  cosmic neutrinos}'',
  \href{http://dx.doi.org/10.1088/1475-7516/2014/06/049}{\blue JCAP {\bfseries
  1406} (2014) 049},
\href{http://arxiv.org/abs/1402.6523}{{\blue arXiv:1402.6523 [hep-ph]}}.

\bibitem{Dias:2014osa}
A.~Dias, A.~Machado, C.~Nishi, A.~Ringwald, and P.~Vaudrevange, ``{The Quest
  for an Intermediate-Scale Accidental Axion and Further ALPs}'',
  \href{http://dx.doi.org/10.1007/JHEP06(2014)037}{\blue JHEP {\bfseries 1406}
  (2014) 037},
\href{http://arxiv.org/abs/1403.5760}{{\blue arXiv:1403.5760 [hep-ph]}}.

\bibitem{Ahn:2014gva}
Y.~Ahn, ``{Flavored Peccei-Quinn symmetry}'',
\href{http://arxiv.org/abs/1410.1634}{{\blue arXiv:1410.1634 [hep-ph]}}.

\bibitem{Ma:2014yka}
E.~Ma, ``{Syndetic Model of Fundamental Interactions}'',
\href{http://arxiv.org/abs/1411.6679}{{\blue arXiv:1411.6679 [hep-ph]}}.

\bibitem{Kim:1979if}
J.~E. Kim, ``{Weak Interaction Singlet and Strong CP Invariance}'',
\href{http://dx.doi.org/10.1103/PhysRevLett.43.103}{\blue Phys.Rev.Lett.
  {\bfseries 43} (1979) 103}.

\bibitem{Shifman:1979if}
M.~A. Shifman, A.~Vainshtein, and V.~I. Zakharov, ``{Can Confinement Ensure
  Natural CP Invariance of Strong Interactions?}'',
\href{http://dx.doi.org/10.1016/0550-3213(80)90209-6}{\blue Nucl.Phys.
  {\bfseries B166} (1980) 493}.

\bibitem{Zhitnitsky:1980tq}
A.~Zhitnitsky, ``{On Possible Suppression of the Axion Hadron Interactions. (In
  Russian)}'',
Sov.J.Nucl.Phys. {\bfseries 31} (1980) 260.

\bibitem{Dine:1981rt}
M.~Dine, W.~Fischler, and M.~Srednicki, ``{A Simple Solution to the Strong CP
  Problem with a Harmless Axion}'',
\href{http://dx.doi.org/10.1016/0370-2693(81)90590-6}{\blue Phys.Lett.
  {\bfseries B104} (1981) 199}.

\bibitem{Jaeckel:2013uva}
J.~Jaeckel, ``{A Family of WISPy Dark Matter Candidates}'',
  \href{http://dx.doi.org/10.1016/j.physletb.2014.03.005}{\blue Phys.Lett.
  {\bfseries B732} (2014) 1--7},
\href{http://arxiv.org/abs/1311.0880}{{\blue arXiv:1311.0880 [hep-ph]}}.

\bibitem{Morrissey:2012db}
D.~E. Morrissey and M.~J. Ramsey-Musolf, ``{Electroweak baryogenesis}'',
  \href{http://dx.doi.org/10.1088/1367-2630/14/12/125003}{\blue New J.Phys.
  {\bfseries 14} (2012) 125003},
\href{http://arxiv.org/abs/1206.2942}{{\blue arXiv:1206.2942 [hep-ph]}}.

\bibitem{Srednicki:1985xd}
M.~Srednicki, ``{Axion Couplings to Matter. 1. CP Conserving Parts}'',
\href{http://dx.doi.org/10.1016/0550-3213(85)90054-9}{\blue Nucl.Phys.
  {\bfseries B260} (1985) 689}.

\bibitem{Arhrib:2011uy}
A.~Arhrib, R.~Benbrik, M.~Chabab, G.~Moultaka, M.~Peyranere, {\em et~al.},
  ``{The Higgs Potential in the Type II Seesaw Model}'',
  \href{http://dx.doi.org/10.1103/PhysRevD.84.095005}{\blue Phys.Rev.
  {\bfseries D84} (2011) 095005},
\href{http://arxiv.org/abs/1105.1925}{{\blue arXiv:1105.1925 [hep-ph]}}.

\bibitem{McDonald:2003zj}
K.~L. McDonald and B.~McKellar, ``{Evaluating the two loop diagram responsible
  for neutrino mass in Babu's model}'',
\href{http://arxiv.org/abs/hep-ph/0309270}{{\blue arXiv:hep-ph/0309270
  [hep-ph]}}.

\bibitem{Bertolini:1987kz}
S.~Bertolini and A.~Santamaria, ``{The Doublet Majoron Model and Solar Neutrino
  Oscillations}'',
\href{http://dx.doi.org/10.1016/0550-3213(88)90100-9}{\blue Nucl.Phys.
  {\bfseries B310} (1988) 714}.

\bibitem{Kim:1986ax}
J.~E. Kim, ``{Light Pseudoscalars, Particle Physics and Cosmology}'',
\href{http://dx.doi.org/10.1016/0370-1573(87)90017-2}{\blue Phys.Rept.
  {\bfseries 150} (1987) 1--177}.

\bibitem{Volkas:1988cm}
R.~Volkas, A.~Davies, and G.~C. Joshi, ``{Naturalness of the invisible axion
  model}'',
\href{http://dx.doi.org/10.1016/0370-2693(88)91084-2}{\blue Phys.Lett.
  {\bfseries B215} (1988) 133}.

\bibitem{Foot:2013hna}
R.~Foot, A.~Kobakhidze, K.~L. McDonald, and R.~R. Volkas, ``{Poincare
  Protection for a Natural Electroweak Scale}'',
  \href{http://dx.doi.org/10.1103/PhysRevD.89.115018}{\blue Phys.Rev.
  {\bfseries D89} (2014) 115018},
\href{http://arxiv.org/abs/1310.0223}{{\blue arXiv:1310.0223 [hep-ph]}}.

\bibitem{deGouvea:2014xba}
A.~de~Gouvea, D.~Hernandez, and T.~M.~P. Tait, ``{Criteria for Natural
  Hierarchies}'', \href{http://dx.doi.org/10.1103/PhysRevD.89.115005}{\blue
  Phys.Rev. {\bfseries D89} (2014) 115005},
\href{http://arxiv.org/abs/1402.2658}{{\blue arXiv:1402.2658 [hep-ph]}}.

\bibitem{Giudice:2014tma}
G.~F. Giudice, G.~Isidori, A.~Salvio, and A.~Strumia, ``{Softened Gravity and
  the Extension of the Standard Model up to Infinite Energy}'',
\href{http://arxiv.org/abs/1412.2769}{{\blue arXiv:1412.2769 [hep-ph]}}.

\bibitem{Holdom:2014hla}
B.~Holdom, J.~Ren, and C.~Zhang, ``{Stable Asymptotically Free Extensions
  (SAFEs) of the Standard Model}'',
  \href{http://dx.doi.org/10.1007/JHEP03(2015)028}{\blue JHEP {\bfseries 1503}
  (2015) 028},
\href{http://arxiv.org/abs/1412.5540}{{\blue arXiv:1412.5540 [hep-ph]}}.

\bibitem{Kamionkowski:1992mf}
M.~Kamionkowski and J.~March-Russell, ``{Planck scale physics and the
  Peccei-Quinn mechanism}'',
  \href{http://dx.doi.org/10.1016/0370-2693(92)90492-M}{\blue Phys.Lett.
  {\bfseries B282} (1992) 137--141},
\href{http://arxiv.org/abs/hep-th/9202003}{{\blue arXiv:hep-th/9202003
  [hep-th]}}.

\bibitem{Kallosh:1995hi}
R.~Kallosh, A.~D. Linde, D.~A. Linde, and L.~Susskind, ``{Gravity and global
  symmetries}'', \href{http://dx.doi.org/10.1103/PhysRevD.52.912}{\blue
  Phys.Rev. {\bfseries D52} (1995) 912--935},
\href{http://arxiv.org/abs/hep-th/9502069}{{\blue arXiv:hep-th/9502069
  [hep-th]}}.

\bibitem{Holman:1992us}
R.~Holman, S.~D. Hsu, T.~W. Kephart, E.~W. Kolb, R.~Watkins, {\em et~al.},
  ``{Solutions to the strong CP problem in a world with gravity}'',
  \href{http://dx.doi.org/10.1016/0370-2693(92)90491-L}{\blue Phys.Lett.
  {\bfseries B282} (1992) 132--136},
\href{http://arxiv.org/abs/hep-ph/9203206}{{\blue arXiv:hep-ph/9203206
  [hep-ph]}}.

\bibitem{Dvali:2013cpa}
G.~Dvali, S.~Folkerts, and A.~Franca, ``{On How Neutrino Protects the Axion}'',
  \href{http://dx.doi.org/10.1103/PhysRevD.89.105025}{\blue Phys.Rev.
  {\bfseries D89} (2014) 105025},
\href{http://arxiv.org/abs/1312.7273}{{\blue arXiv:1312.7273 [hep-th]}}.

\bibitem{Lebedev:2012zw}
O.~Lebedev, ``{On Stability of the Electroweak Vacuum and the Higgs Portal}'',
  \href{http://dx.doi.org/10.1140/epjc/s10052-012-2058-2}{\blue Eur.Phys.J.
  {\bfseries C72} (2012) 2058},
\href{http://arxiv.org/abs/1203.0156}{{\blue arXiv:1203.0156 [hep-ph]}}.

\bibitem{Chao:2012mx}
W.~Chao, M.~Gonderinger, and M.~J. Ramsey-Musolf, ``{Higgs Vacuum Stability,
  Neutrino Mass, and Dark Matter}'',
  \href{http://dx.doi.org/10.1103/PhysRevD.86.113017}{\blue Phys.Rev.
  {\bfseries D86} (2012) 113017},
\href{http://arxiv.org/abs/1210.0491}{{\blue arXiv:1210.0491 [hep-ph]}}.

\bibitem{Dev:2013ff}
P.~Bhupal~Dev, D.~K. Ghosh, N.~Okada, and I.~Saha, ``{125 GeV Higgs Boson and
  the Type-II Seesaw Model}'', \href{http://dx.doi.org/10.1007/JHEP03(2013)150,
  10.1007/JHEP05(2013)049, 10.1007/JHEP05(2013)049,
  10.1007/JHEP03(2013)150}{\blue JHEP {\bfseries 1303} (2013) 150},
\href{http://arxiv.org/abs/1301.3453}{{\blue arXiv:1301.3453 [hep-ph]}}.

\bibitem{Kobakhidze:2013pya}
A.~Kobakhidze and A.~Spencer-Smith, ``{Neutrino Masses and Higgs Vacuum
  Stability}'', \href{http://dx.doi.org/10.1007/JHEP08(2013)036}{\blue JHEP
  {\bfseries 1308} (2013) 036},
\href{http://arxiv.org/abs/1305.7283}{{\blue arXiv:1305.7283 [hep-ph]}}.

\bibitem{DiLuzio:2014bua}
L.~Di~Luzio and L.~Mihaila, ``{On the gauge dependence of the Standard Model
  vacuum instability scale}'',
  \href{http://dx.doi.org/10.1007/JHEP06(2014)079}{\blue JHEP {\bfseries 1406}
  (2014) 079},
\href{http://arxiv.org/abs/1404.7450}{{\blue arXiv:1404.7450 [hep-ph]}}.

\bibitem{Andreassen:2014gha}
A.~Andreassen, W.~Frost, and M.~D. Schwartz, ``{Consistent Use of the Standard
  Model Effective Potential}'',
\href{http://arxiv.org/abs/1408.0292}{{\blue arXiv:1408.0292 [hep-ph]}}.

\bibitem{Espinosa:2007qp}
J.~Espinosa, G.~Giudice, and A.~Riotto, ``{Cosmological implications of the
  Higgs mass measurement}'',
  \href{http://dx.doi.org/10.1088/1475-7516/2008/05/002}{\blue JCAP {\bfseries
  0805} (2008) 002},
\href{http://arxiv.org/abs/0710.2484}{{\blue arXiv:0710.2484 [hep-ph]}}.

\bibitem{Gunion:2002zf}
J.~F. Gunion and H.~E. Haber, ``{The CP conserving two Higgs doublet model: The
  Approach to the decoupling limit}'',
  \href{http://dx.doi.org/10.1103/PhysRevD.67.075019}{\blue Phys.Rev.
  {\bfseries D67} (2003) 075019},
\href{http://arxiv.org/abs/hep-ph/0207010}{{\blue arXiv:hep-ph/0207010
  [hep-ph]}}.

\bibitem{Akeroyd:2009nu}
A.~Akeroyd, M.~Aoki, and H.~Sugiyama, ``{Lepton Flavour Violating Decays
  $\tau\to \bar{l}ll$ and $\mu\to e \gamma$ in the Higgs Triplet Model}'',
  \href{http://dx.doi.org/10.1103/PhysRevD.79.113010}{\blue Phys.Rev.
  {\bfseries D79} (2009) 113010},
\href{http://arxiv.org/abs/0904.3640}{{\blue arXiv:0904.3640 [hep-ph]}}.

\bibitem{Dumont:2014wha}
B.~Dumont, J.~F. Gunion, Y.~Jiang, and S.~Kraml, ``{Constraints on and future
  prospects for Two-Higgs-Doublet Models in light of the LHC Higgs signal}'',
  \href{http://dx.doi.org/10.1103/PhysRevD.90.035021}{\blue Phys.Rev.
  {\bfseries D90} (2014) 035021},
\href{http://arxiv.org/abs/1405.3584}{{\blue arXiv:1405.3584 [hep-ph]}}.

\bibitem{Chun:2012jw}
E.~J. Chun, H.~M. Lee, and P.~Sharma, ``{Vacuum Stability, Perturbativity, EWPD
  and Higgs-to-diphoton rate in Type II Seesaw Models}'',
  \href{http://dx.doi.org/10.1007/JHEP11(2012)106}{\blue JHEP {\bfseries 1211}
  (2012) 106},
\href{http://arxiv.org/abs/1209.1303}{{\blue arXiv:1209.1303 [hep-ph]}}.

\bibitem{ATLAS:2012hi}
{ ATLAS Collaboration}, G.~Aad {\em et~al.}, ``{Search for doubly-charged Higgs
  bosons in like-sign dilepton final states at $\sqrt{s}=7$ TeV with the ATLAS
  detector}'', \href{http://dx.doi.org/10.1140/epjc/s10052-012-2244-2}{\blue
  Eur.Phys.J. {\bfseries C72} (2012) 2244},
\href{http://arxiv.org/abs/1210.5070}{{\blue arXiv:1210.5070 [hep-ex]}}.

\bibitem{Chatrchyan:2012ya}
{ CMS Collaboration}, S.~Chatrchyan {\em et~al.}, ``{A search for a
  doubly-charged Higgs boson in $pp$ collisions at $\sqrt{s}=7$ TeV}'',
  \href{http://dx.doi.org/10.1140/epjc/s10052-012-2189-5}{\blue Eur.Phys.J.
  {\bfseries C72} (2012) 2189},
\href{http://arxiv.org/abs/1207.2666}{{\blue arXiv:1207.2666 [hep-ex]}}.

\bibitem{Melfo:2011nx}
A.~Melfo, M.~Nemevsek, F.~Nesti, G.~Senjanovic, and Y.~Zhang, ``{Type II Seesaw
  at LHC: The Roadmap}'',
  \href{http://dx.doi.org/10.1103/PhysRevD.85.055018}{\blue Phys.Rev.
  {\bfseries D85} (2012) 055018},
\href{http://arxiv.org/abs/1108.4416}{{\blue arXiv:1108.4416 [hep-ph]}}.

\bibitem{Chen:2014qda}
C.-H. Chen and T.~Nomura, ``{Search for $\delta^{\pm\pm}$ with new decay
  patterns at LHC}'',
\href{http://arxiv.org/abs/1411.6412}{{\blue arXiv:1411.6412 [hep-ph]}}.

\bibitem{Profumo:2007wc}
S.~Profumo, M.~J. Ramsey-Musolf, and G.~Shaughnessy, ``{Singlet Higgs
  phenomenology and the electroweak phase transition}'',
  \href{http://dx.doi.org/10.1088/1126-6708/2007/08/010}{\blue JHEP {\bfseries
  0708} (2007) 010},
\href{http://arxiv.org/abs/0705.2425}{{\blue arXiv:0705.2425 [hep-ph]}}.

\bibitem{AbdusSalam:2013eya}
S.~S. AbdusSalam and T.~A. Chowdhury, ``{Scalar Representations in the Light of
  Electroweak Phase Transition and Cold Dark Matter Phenomenology}'',
  \href{http://dx.doi.org/10.1088/1475-7516/2014/05/026}{\blue JCAP {\bfseries
  1405} (2014) 026},
\href{http://arxiv.org/abs/1310.8152}{{\blue arXiv:1310.8152 [hep-ph]}}.

\bibitem{Chowdhury:2011ga}
T.~A. Chowdhury, M.~Nemevsek, G.~Senjanovic, and Y.~Zhang, ``{Dark Matter as
  the Trigger of Strong Electroweak Phase Transition}'',
  \href{http://dx.doi.org/10.1088/1475-7516/2012/02/029}{\blue JCAP {\bfseries
  1202} (2012) 029},
\href{http://arxiv.org/abs/1110.5334}{{\blue arXiv:1110.5334 [hep-ph]}}.

\bibitem{Haber:2012np}
H.~E. Haber and Z.~Surujon, ``{A Group-theoretic Condition for Spontaneous CP
  Violation}'', \href{http://dx.doi.org/10.1103/PhysRevD.86.075007}{\blue
  Phys.Rev. {\bfseries D86} (2012) 075007},
\href{http://arxiv.org/abs/1201.1730}{{\blue arXiv:1201.1730 [hep-ph]}}.

\bibitem{Geng:1988ty}
C.~Geng, X.~Jiang, and J.~N. Ng, ``{A Minimal Model of Spontaneous $T$
  Violation With an Invisible Axion}'',
\href{http://dx.doi.org/10.1103/PhysRevD.38.1628}{\blue Phys.Rev. {\bfseries
  D38} (1988) 1628}.

\bibitem{Servant:2014bla}
G.~Servant, ``{Baryogenesis from Strong $CP$ Violation and the QCD Axion}'',
  \href{http://dx.doi.org/10.1103/PhysRevLett.113.171803}{\blue Phys.Rev.Lett.
  {\bfseries 113} no.~17, (2014) 171803},
\href{http://arxiv.org/abs/1407.0030}{{\blue arXiv:1407.0030 [hep-ph]}}.

\bibitem{Allison:2014zya}
K.~Allison, C.~T. Hill, and G.~G. Ross, ``{Ultra-weak sector, Higgs boson mass,
  and the dilaton}'',
  \href{http://dx.doi.org/10.1016/j.physletb.2014.09.041}{\blue Phys.Lett.
  {\bfseries B738} (2014) 191--195},
\href{http://arxiv.org/abs/1404.6268}{{\blue arXiv:1404.6268 [hep-ph]}}.

\bibitem{Allison:2014hna}
K.~Allison, C.~T. Hill, and G.~G. Ross, ``{An ultra-weak sector, the strong CP
  problem and the pseudo-Goldstone dilaton}'',
\href{http://arxiv.org/abs/1409.4029}{{\blue arXiv:1409.4029 [hep-ph]}}.

\end{thebibliography}\endgroup

\end{document}